\documentclass[epj]{svjour}
\usepackage[dvips]{graphicx}
\usepackage{amsmath}

\usepackage{amstext}

\usepackage{graphics}
\usepackage{exscale}
\usepackage{epsfig}
\usepackage{changebar}
\usepackage[latin1]{inputenc}
\usepackage{bm}
\usepackage{bbm}

\usepackage{rotating}
\usepackage{latexsym}

\usepackage{psfrag}

\renewcommand{\epsilon}{\varepsilon}

\newcommand{\integral}[3]{\!\int\limits_{#2}^{#3}\!\!{\rm d}#1\;}

\newcommand{\elcre}[2]{ c^{\dagger}_{#1,#2}}
\newcommand{\elann}[2]{ c^{}_{#1,#2}}

\newcommand{\e}{\mathrm e}

\newcommand{\vct}[1]{\bm #1}
\newcommand{\vk}{{\bm k}}

\newcommand{\vkF}{\vk_{{\scriptscriptstyle \mathrm{F}}}}
\newcommand{\epsF}{\epsilon_{{\scriptscriptstyle \mathrm{F}}}}

\newcommand{\ord}[1]{{\cal O}(#1)}

\newcommand{\thGf}{{\cal G}}

\newcommand{\Imag}{\mathrm{Im}}

\newcommand{\hc}{\mathrm{h.c.}}

\begin{document}

\title{Renormalized quasiparticles in antiferromagnetic states of the Hubbard model}
\author{J. Bauer and A.C. Hewson}
\institute{Department of Mathematics, Imperial College, London SW7 2AZ,
  United Kingdom}
\date{\today} 
\abstract{
We analyze the properties of the quasiparticle excitations of metallic
antiferromagnetic states in a strongly correlated electron system. The study is based on 
dynamical mean field theory (DMFT) for the infinite dimensional Hubbard model with
antiferromagnetic symmetry breaking. Self-consistent solutions of the DMFT
equations are calculated using the  numerical renormalization group (NRG). The
low energy behavior in these results is then analyzed in terms 
of renormalized quasiparticles. The parameters for these quasiparticles are
calculated directly from the NRG derived self-energy, and also from the low
energy fixed point of the effective impurity model. From these the quasiparticle
weight and the effective mass are deduced. We show that
the main low energy features of the $\vk$-resolved spectral density can be understood in
terms of the quasiparticle picture. We also  find that Luttinger's theorem
is satisfied for the total  electron number in the doped  antiferromagnetic
state. }

\maketitle

\section{Introduction}

The nature of the metallic antiferromagnetically ordered state in strongly
correlated systems has been subject of study for over two decades,
but still remains to be fully understood. Interest in this topic
has been stimulated by the fact that the high temperature superconductivity of
the cuprates emerges from the doping  of an  antiferromagnetic insulating
compound, such as La$_2$CuO$_4$ \cite{And87,LNW06}. The simplest models to describe
the electrons in the CuO$_2$ planes of the cuprates are the two dimensional
$t$-$J$-model and  Hubbard model. Much of the initial effort went into the study
of a single hole state of these models in an antiferromagnetic background. For
the motion of this hole, there is a competition between the gain in kinetic
energy from the hopping and its disruptive effect on the antiferromagnetic
order, and consequent loss of potential energy. As a result a hole excitation becomes a
quasiparticle or magnetic polaron, heavily dressed by antiferromagnetic spin
fluctuations (see review article by Dagotto \cite{Dag94} and references therein).
Much of this work, however,  relied on exact diagonalization or quantum Monte
Carlo methods, which are limited to small clusters and very few hole
excitations, and cannot be readily extended to study the many-hole, finite
doping situation. 


More recent studies capable of describing finite doping  have
concentrated on the relation between the  antiferromagnetic fluctuations and
superconducting order (for a review see \cite{TKS06} and the references therein).
One of the main motivations is to understand whether the exchange of these types 
of fluctuations can provide a purely electronic mechanism for inducing
superconductivity. Here, in this paper, we focus on  the metallic
antiferromagnetism, the doped state with long range antiferromagnetic order. 
Our interest is to examine how well the low energy excitations in this ordered 
state can be described in terms  renormalized quasiparticles. To tackle this
problem we use the infinite dimensional Hubbard model.

The simplification in the infinite dimensional  limit is  that the electron
self-energy becomes local in character, with no wavevector
dependence \cite{MV89,Mue89}.
The self-energy then depends only on the frequency, as is the case for
impurity models, allowing the lattice problem to be cast in the form of a
self-consistent impurity model. There are several reasonably accurate
techniques for solving this effective impurity problem, a very accurate one
for the zero and low temperature regime being the numerical renormalization
group approach (NRG) \cite{BCP07pre}.

Recently we studied the effect of a magnetic field on the quasiparticle
excitations in the strong correlation regime of the infinite dimensional
Hubbard model using the NRG method \cite{BH07bpre}. We also extended a form of renormalized
perturbation theory (RPT), originally developed for impurity models
\cite{Hew93}, to this model, and used it to calculate the local dynamic spin
susceptibilities, obtaining results  in good overall  agreement with those
from  the NRG. In this paper we extend this combination of renormalization
techniques, NRG and RPT within dynamical mean field theory, to look at the low
energy excitations of the infinite dimensional
Hubbard model in a staggered field, and in antiferromagnetic broken symmetry
states. Extensive calculations of the antiferromagnetic states in the Hubbard model
using the DMFT-NRG approach have already been reported in the paper of
Zitzler, Pruschke and Bulla \cite{ZPB02}. We confirm their results for the
phase diagram and extend the calculations and analysis to the description of
the low energy renormalized excitations, and how these can be described within
the framework of a renormalized perturbation theory. 
 
\section{Antiferromagnetic Broken Symmetry in DMFT}

In considering the response of the Hubbard model \cite{Hub63} to a staggered
magnetic field and antiferromagnetic order, we take the case of a bipartite
lattice, which consists of two sublattices $A$ and $B$ such that the nearest
neighbors of a site in the $A$ sublattice are on the $B$ sublattice and vice
versa.  The Hamiltonian 
for the Hubbard model can be written in the form,
\begin{eqnarray}
H_{\mu}&=&
\sum_{i,j,\sigma}(t_{ij}\elcre{A,i}{\sigma}\elann{B,j}{\sigma}+\hc) 
+U\sum_{i,\alpha} n_{\alpha,i,\uparrow}n_{\alpha,i,\downarrow} \nonumber \\
&&-\sum_{i,\sigma}(\mu_{\sigma}\elcre{A,i}{\sigma}\elann{A,i}{\sigma}
+\mu_{-\sigma}\elcre{B,i}{\sigma}\elann{B,i}{\sigma}),
\end{eqnarray}
where the hopping matrix element is taken as $t_{ij}=-t$ between nearest sites $i$
and $j$ only, and zero otherwise, and $\alpha=A,B$. A staggered field $H^{i}_{\rm s}$
\begin{equation}
  H^{i}_{\rm s}=
\left\{
\begin{array} {rr}
 H & {\rm for}\; i\in A\;{\rm sublattice} \\
-H & {\rm for}\; i\in B\;{\rm sublattice}
\end{array}
\right.
\end{equation}
has been included so that $\mu_\sigma=\mu+\sigma h$,
where $h=g \mu_{\rm B}H/2$ with the Bohr magneton $\mu_{\rm B}$. The
non-interacting part of the Hamiltonian $H_{0,\mu}$  can be diagonalized in
terms of Bloch states and then expressed in the form, 
\begin{equation}
H_{0,\mu}=
\sum_{\vk,\sigma}C^{\dagger}_{\vk,\sigma}M^{}_{\vk,\sigma}C^{}_{\vk,\sigma}.
\end{equation}
 where
$C_{\vk,\sigma}^{\dagger}=(\elcre{A,\vk}{\sigma},\elcre{B,\vk}{\sigma})$,
and 
the matrix  $M_{\vk,\sigma}$  is given by
\begin{equation}
M_{\vk,\sigma}=
\left(
\begin{array} {cc}
-\mu_{\sigma} & \epsilon_{\vk} \\
\epsilon_{\vk} & -\mu_{-\sigma} 
\end{array}
\right).
\end{equation}
The $\vk$ sums run over a reduced Brillouin zone,
and the energy of the Bloch state is $\epsilon_{\vk}=\sum_{j}t_{ij}\e^{i{
    (\vct R_i-\vct R_j)}\cdot{\vk}}$.
The free Green's function matrix $\underline{G}^0_{\vk,\sigma}(\omega)$ is  given by 
$(\omega-M_{\vk,\sigma})^{-1}$. The poles of the free Green's function give
the elementary single particle excitations, which are given by
\begin{equation}
  E_{\vk,\pm}^0(U=0)=-\mu_0(h)\pm\sqrt{h^2+\epsilon_{\vk}^2},
\label{freeexcit}
\end{equation}
where $\mu_0(h)$ is the chemical potential of the noninteracting system in a
staggered field.
This illustrates that the electronic excitations are split into two subbands
for a finite staggered field.

Notice that we have adopted a special choice of basis 
$\{\elann{A,\vk}{\sigma},\elann{B,\vk}{\sigma}\}$ here \cite{GKKR96,ZPB02}.
Another common basis to study antiferromagnetic and spin density wave symmetry (SDW)
breaking is $\{\elann{\vk}{\sigma},\elann{\vk+\vct q_0}{\sigma}\}$, where
$\vct q_0$ is the reciprocal lattice vector for commensurate SDW ordering. 
The bases can be  related by a linear transformation,
\begin{equation}
\left(
\begin{array} {c}
\!\!\elann{\vk}{\sigma} \!\\
\!\!\elann{\vk+\vct q_0}{\sigma}\!
\end{array}
\right)
=
\frac1{\sqrt2}
\left(
\begin{array} {rr}
 1 & -1 \\
 1 &  1 
\end{array}
\right)
\left(
\begin{array} {c}
\!\!\elann{A,\vk}{\sigma}\! \\
\!\!\elann{B,\vk}{\sigma}\!
\end{array}
\right).
\label{basisrelation}
\end{equation}
For the latter basis the matrix $M_{\vk,\sigma}$ would be diagonal in the
kinetic energy term and the symmetry breaking is offdiagonal.
For our study in the DMFT framework the $A-B$-sublattice basis is, however, more 
convenient and we will use it throughout the rest of this paper. It is
possible, of course, to relate the obtained quantities with the help of
(\ref{basisrelation}) to the  $\{\elann{\vk}{\sigma},\elann{\vk+\vct
  q_0}{\sigma}\}$ basis.

We can generalize the equations to the interacting problem by introducing
a self-energy $\Sigma_{\alpha,\vk,\sigma}(\omega)$, so that the matrix Green's
function can be written in the form    
\begin{equation}
\underline{G}_{\vk,\sigma}(\omega) \!=\!
\frac1{\zeta_{A,\vk,\sigma}(\omega)\zeta_{B,\vk,\sigma}(\omega) -\epsilon_{\vk}^2}
\! \left(\!\!\!
\begin{array} {cc}
 \zeta_{B,\vk,\sigma}(\omega) & -\epsilon_{\vk} \\
-\epsilon_{\vk} & \zeta_{A,\vk,\sigma}(\omega)
\end{array}
\!\!\!\right),
\label{kgf}
\end{equation}
where
$\zeta_{\alpha,\vk,\sigma}(\omega)=\omega+\mu_{\sigma}-\Sigma_{\alpha,\vk,\sigma}(\omega)$.
As we are dealing with the infinite dimensional limit of the model,  
we take the self-energy to be local so we can drop the $\vk$ index. This is
the reason why the self-energy has a single site index $\alpha=A,B$ and no
offdiagonal terms appear in equation (\ref{kgf}).
The symmetry of the bipartite lattice gives
$\Sigma_{B,\sigma}(\omega)=\Sigma_{A,-\sigma}(\omega)\equiv
\Sigma_{-\sigma}(\omega)$ and hence
\[
\zeta_{B,-\sigma}(\omega)=\zeta_{A,\sigma}(\omega)\equiv\zeta_{\sigma}(\omega),
\]
where we have simplified the notation. To determine these
quantities $\Sigma_{\sigma}(\omega)$ it is sufficient to focus on the $A$
sublattice only. 

Summing the first component in the Green's function in equation (\ref{kgf})
over ${\vk}$ we obtain the Green's function for a site on the $A$ sublattice,
$G_{\sigma}^{\mathrm{loc}}(\omega)$, 
\begin{equation}
 G_{\sigma}^{\mathrm{loc}}(\omega)
=\frac{\zeta_{-\sigma}(\omega)}{\sqrt{\zeta_{\sigma}(\omega)\zeta_{-\sigma}(\omega)}}
\integral{\epsilon}{}{}\frac{\rho_0(\epsilon)}
{\sqrt{\zeta_{\sigma}(\omega)\zeta_{-\sigma}(\omega)}-\epsilon},
\label{afmgfct}
\end{equation}
where $\rho_0(\epsilon)$ is the density of states of the non-interacting
system in the absence of the staggered field.

In the DMFT  this local  Green's function, and the self-energy
$\Sigma_\sigma(\omega)$, are identified with the corresponding quantities for 
an effective impurity model \cite{GKKR96}. This implies that the Green's function
$\thGf_{0,\sigma}(\omega)$  for the effective 
impurity  in the absence of an interaction at the impurity site is given by
\begin{equation}
  \thGf_{0,\sigma}^{-1}(\omega)=G_{\sigma}^{\mathrm{loc}}(\omega)^{-1}
  +\Sigma_{\sigma}(\omega).
\label{afmselfen}
\end{equation}
We can take the form of this impurity model to correspond to an
Anderson model \cite{And61} in a magnetic field,
\begin{eqnarray}
&&H_{\rm AM}=\sum\sb {\sigma}\epsilon\sb {\mathrm{d},\sigma} d\sp {\dagger}\sb
{\sigma}  
d\sp {}\sb {\sigma}+Un\sb {\mathrm{d},\uparrow}n\sb {\mathrm{d},\downarrow} \label{ham}\\
&& +\sum\sb {{ k},\sigma}( V\sb {\vk,\sigma}d\sp {\dagger}\sb {\sigma}
c\sp {}\sb {{ k},\sigma}+ V\sb { \vk,\sigma}\sp *c\sp {\dagger}\sb {{
k},\sigma}d\sp {}\sb {\sigma})+\sum_{\vk,\sigma}\epsilon\sb {\vk,\sigma}c\sp
{\dagger}\sb {{\vk},\sigma} c_{\vk,\sigma}, \nonumber
\end{eqnarray}
where $\epsilon_{\mathrm{d},\sigma}=\epsilon_{\rm d}-\sigma h$
is the energy of the localized  level at an impurity site   in a magnetic
field $H$, $U$ the interaction at this local site, and $V_{\vk,\sigma}$ the
hybridization matrix element to a band of conduction electrons of spin
$\sigma$ with energy $\epsilon_{\vk,\sigma}$. As we are focusing on an $A$ site
as the impurity we take $H=H_s$. 

The one-electron Green's function for the impurity site of this model is given by
\begin{equation}
G^{\mathrm{imp}}_{\sigma}(\omega)=\frac{1}
{\omega-\epsilon_{\mathrm{d}\sigma}-K_\sigma(\omega)-\Sigma_\sigma(\omega)},
\label{gfdmft}
\end{equation}
where 
\begin{equation}
K_\sigma(\omega)=\sum_{\vk}\frac{|V_{\vk,\sigma}|^2}{\omega-\epsilon_{\vk,\sigma}}.
\label{lgf}
\end{equation}
If this impurity Green's function is equated to
the local lattice Green's function $G_{\sigma}^{\mathrm{loc}}(\omega)$,
we identify $\epsilon_{\mathrm{d}\sigma}=-\mu_\sigma$ and from equation
 (\ref{afmselfen}), $K_\sigma(\omega)$ is given by  
\begin{equation}
K_{\sigma}(\omega)=\omega+\mu_\sigma-\thGf_{0,\sigma}^{-1}(\omega).
\label{afmmedium}
\end{equation}
The function $K_\sigma(\omega)$ plays the role of the effective medium and has
to be calculated self-consistently.

The self-consistent calculations for $K_\sigma(\omega)$ can usually be
performed  iteratively.  Starting from a conjectured form for
$K_\sigma(\omega)$, the NRG method is used to calculate the self-energy of the
effective Anderson model, from which the impurity Green's function 
$G_{\sigma}^{\mathrm{imp}}(\omega)$ in (\ref{gfdmft}) and the local Green's
function  for the lattice $G_{\sigma}^{\mathrm{loc}}(\omega)$ in
(\ref{afmgfct}) can  be deduced. If these two Green's functions do not agree,
then equation (\ref{afmselfen}) is used to derive a new starting value for
$K_\sigma(\omega)$ and the process continued until self-consistency is
achieved. 

To find  antiferromagnetic solutions, we calculated self-consistent solutions
for a decreasing  sequence of staggered  magnetic fields to see if broken
symmetry solutions of this type exist as the staggered field is reduced to
zero.  For the non-interacting density of states $\rho_0(\epsilon)$  we take
the Gaussian form $\rho_0(\epsilon)=\e^{-(\epsilon/t^*)^2}/\sqrt{\pi}t^*$,
corresponding to an infinite dimensional hypercubic lattice. It is useful to
define an effective  bandwidth $W=2D$ for this density of states  via $D$,
the point at which $\rho_0(D)= \rho_0(0)/\e^2$,  giving $D=\sqrt2 t^*$
corresponding to the choice in reference \cite{Bul99}. In all the results we
present here we take the value $W=4$. 
In the NRG calculations  we have used the improved method \cite{PPA06,WD06pre}
of evaluating the response functions with the complete Anders-Schiller basis
\cite{AS05}, and also determine the self-energy from a higher order Green's
function \cite{BHP98}.

\begin{figure}[!htbp]
\centering
\includegraphics[width=0.45\textwidth]{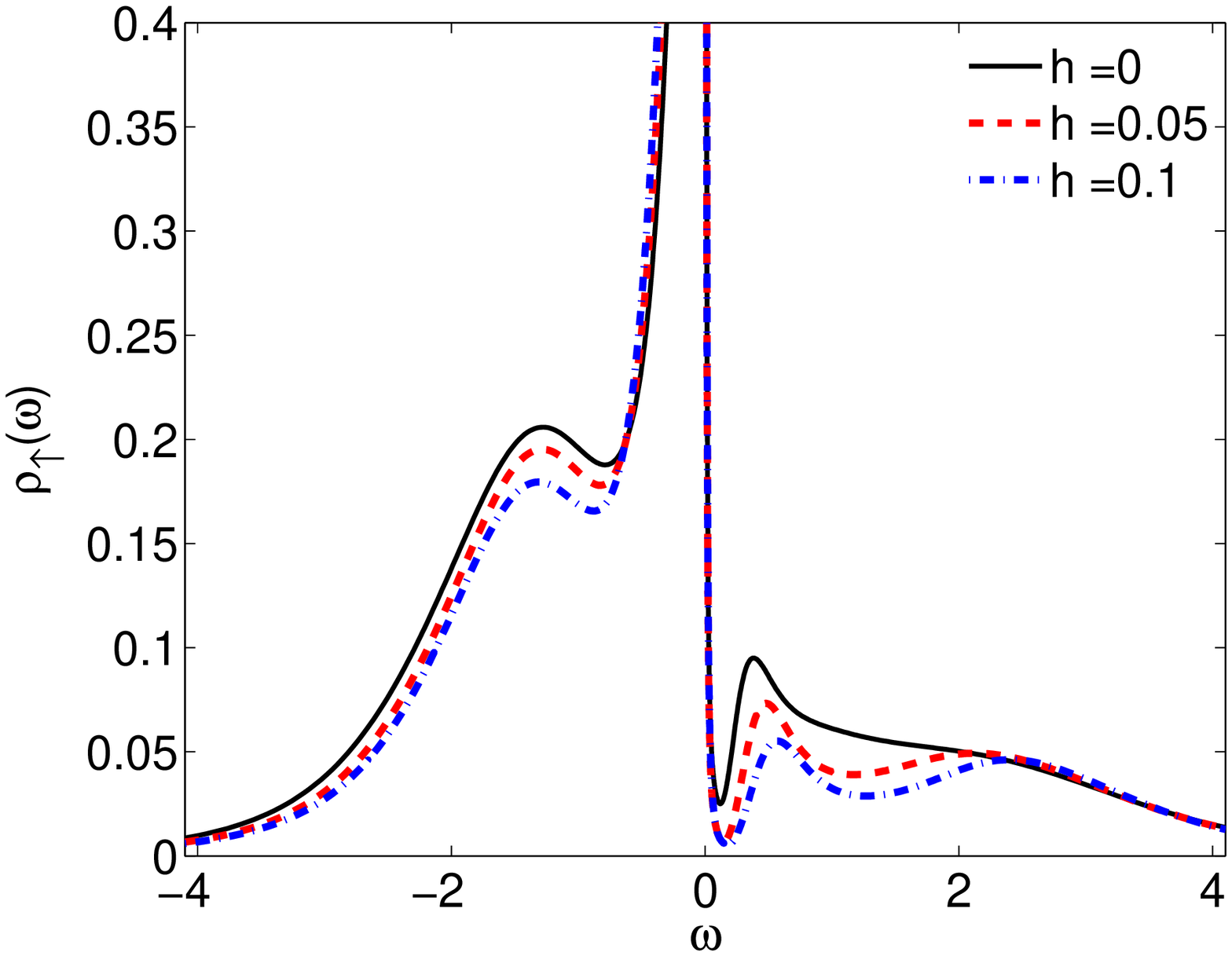}
\includegraphics[width=0.45\textwidth]{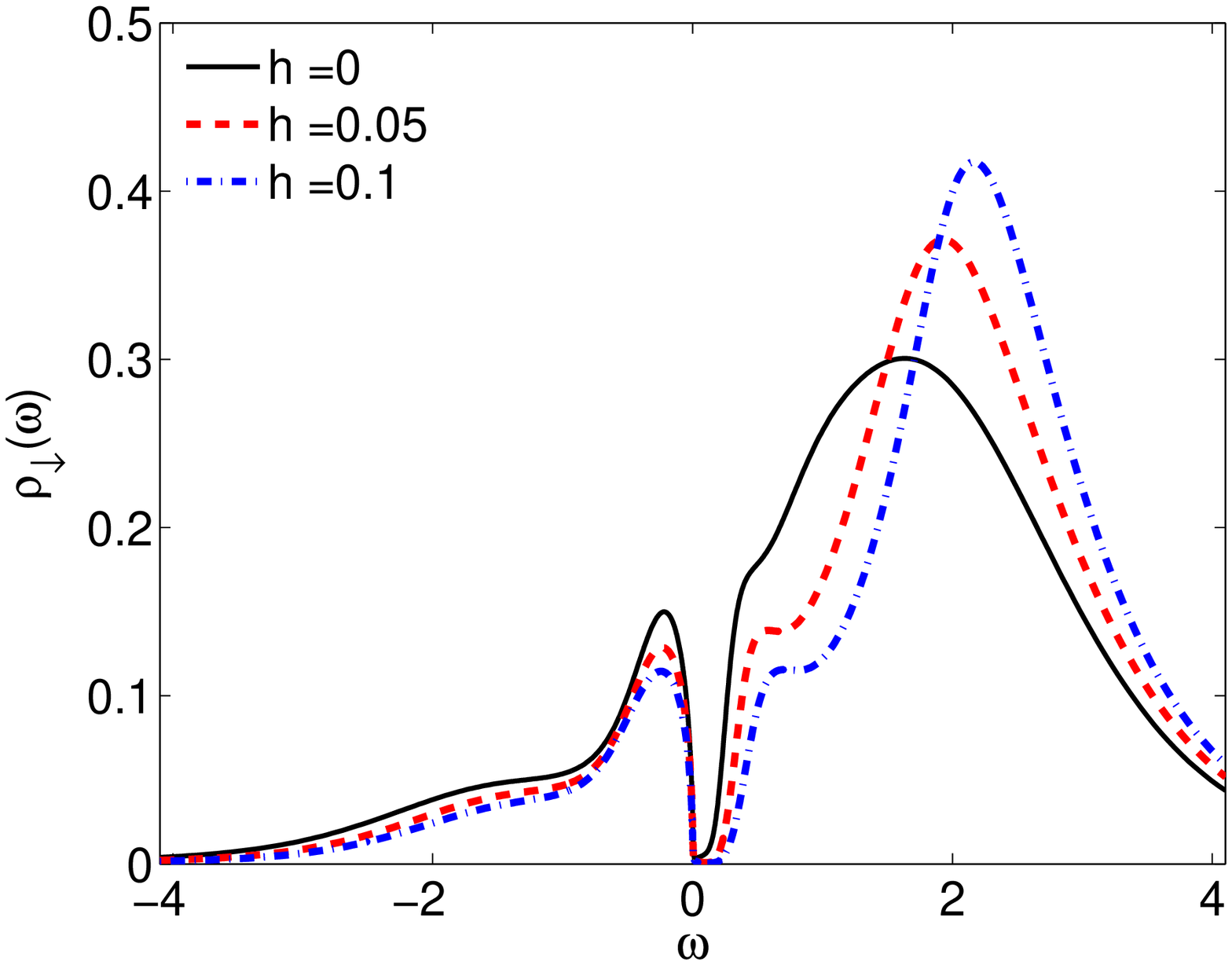}
\caption{(Color online) The spectral densities for the spin-up electrons
  (upper panel) and spin-down electrons (lower panel) at the $A$ site for
  various values of the applied staggered field for $U=3$ and $x=0.95$}
\label{updosdifhU3x0.95}
\end{figure}

In figure \ref{updosdifhU3x0.95} we show the self-consistently 
calculated local spectral density for the spin-up (upper panel) and spin-down
electrons (lower panel) at an $A$ site with $U=3$ and 5\% hole doping (from the
state at  half-filling) for various values of an applied staggered field. 
The staggered magnetic field induces a sublattice magnetization,
\begin{equation}
m_A=\frac12(n_{A,\uparrow}-n_{A,\downarrow}),  
\end{equation}
so that these spectra are quite different. For this set of parameters, this
difference persists as the staggered field is reduced to zero so that we have
a spontaneous sublattice magnetization corresponding to spontaneous 
antiferromagnetic order. For the case away from half filling, $\delta\neq 0$,
we have to keep adjusting the chemical potential when iterating for a
self-consistent solution. It shows a slightly oscillatory behavior when
iterating for a specific filling $x$, and we follow the procedure described in
reference \cite{ZPB02}. This feature is related to the fact that the
calculations are for a metastable ground state and instabilities
to more complicated ground states for antiferromagnetic ordering than the
homogeneous, commensurate N\'eel state, which forms the basis for these DMFT
calculations, can occur \cite{SS89,KMNF90,EKL90,Don95,Don96,Sch90,FJ95b,EKT99,ZPB02}. 
As far as phase separation in the ground state is concerned, the results of
our calculations are generally in line with the conclusions in \cite{ZPB02} as
they are carried out within the same framework. The focus of this work is,
however, the analysis of generic quasiparticle properties in a doped antiferromagnetic
state. We consider the approach as a valid, approximate starting
point for this endeavor, but modifications to the results presented here can occur
for calculations based on a more complicated ground states not accessible
within the DMFT framework. For a more extensive 
discussion of the applicability of the DMFT in this situation we refer to the
earlier work \cite{ZPB02}. 

\begin{figure}[!htbp]
\centering
\includegraphics[width=0.45\textwidth]{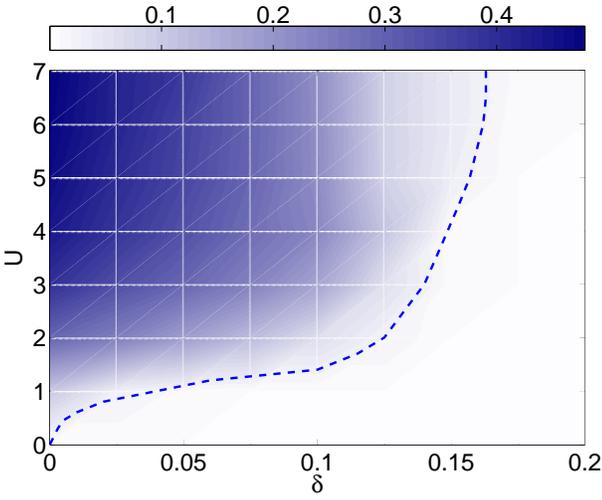}
\caption{(Color online) Phase diagram showing the doping and the $U$
  dependence of the sublattice magnetization $m_A$ as deduced from the
  DMFT-NRG calculations.} 
\label{afmphdia2}
\end{figure}

From results of this type of calculation, we have built up a global
antiferromagnetic/paramagnetic phase diagram as a function of the doping
$\delta$ and the on-site interaction $U$.
This phase diagram is shown in figure  \ref{afmphdia2}, where the value
of the corresponding sublattice magnetization is shown in a false color
plot. We have added a line separating the spontaneously ordered and
paramagnetic regimes.

\noindent
At half filling ($\delta=0$ axis) the spontaneous magnetization increases with
$U$. 
We can see that the antiferromagnetic order from the half filled case persists
when holes are added. The value of the critical doping $\delta_c$ at which the
antiferromagnetism disappears depends on the on-site interaction $U$. We
expect that for small $U$ the critical doping  $\delta_c$ will increase with
$U$ since a tendency to order only appears when an on-site interaction is
present. From the mapping to the $t-J$ model we also expect that for large $U$
the antiferromagnetic coupling $J$ decreases and therefore the order is
destroyed more easily. The values of $U$ are, however, not large enough to
display this trend. 

If we compare these results with the phase diagram given by Zitzler et al.
\cite{ZPB02} we see that they are in very good agreement. In their case
the antiferromagnetic region was mapped out to values of $U\simeq 4.5$.
As the iterations tend to oscillate, as discussed before, there is a  problem
of obtaining a self-consistent antiferromagnetic solution in the large $U$
regime. We have managed to extend the diagram to somewhat 
larger values of $U$ by stabilising the calculations by averaging the
effective medium over a number of iterations.


\section{Local Quasiparticle Parameters}

To examine the nature of the low energy excitations, we will 
assume that the  self-energy $\Sigma_{\sigma}(\omega)$ is non-singular at
$\omega=0$ so that, at least asymptotically,  it can be expanded in powers of
$\omega$. This assumption is not  expected to be valid close to the quantum critical
point when the magnetic order sets in, but to be a reasonable assumption
otherwise. We also assume that the imaginary part of the self-energy
vanishes which is confirmed by the numerical results of the DMFT-NRG calculations.
We will retain terms to order $\omega$ only for the moment. The
higher order corrections will be considered later.  
We then find for $\zeta_{\sigma}(\omega)$,
\begin{eqnarray}
\zeta_{\sigma}(\omega)
  &=&\omega(1-\Sigma'_{\sigma}(0))+\mu_\sigma-\Sigma_{\sigma}(0)      \\
 &=&z_{\sigma}^{-1}(\omega+\tilde\mu_{0,\sigma}),
 \end{eqnarray}
where 
\begin{equation}
\tilde\mu_{0,\sigma}=z_\sigma(\mu-\Sigma_\sigma(0)),\quad
{\rm and}\quad  z_{\sigma}^{-1}=1-\Sigma'_{\sigma}(0).
\label{rplat}
\end{equation}
The interacting Green's function (\ref{kgf}) has poles at the roots of the quadratic
equation, 
\begin{equation} 
\zeta_{\sigma}(\omega)\zeta_{-\sigma}(\omega) -\epsilon_{\vk}^2=0.
\end{equation}
The solutions of this equation are
\begin{equation} 
E^0_{\vk,\pm}=-\tilde\mu\pm \sqrt{\tilde\epsilon_{\vk}^2+
  \Delta\tilde\mu^2},
\label{qpex} 
\end{equation}
where $\tilde\epsilon_{\vk}=\sqrt{z_\uparrow z_\downarrow} \epsilon_{\vk}$, $
\Delta\tilde\mu=(\tilde\mu_{0,\uparrow}-\tilde\mu_{0,\downarrow})/2$, 
and  $ \tilde\mu=(\tilde\mu_{0,\uparrow}+\tilde\mu_{0,\downarrow})/2$.
This has the same form as for the non-interacting system in a staggered
field (\ref{freeexcit}), so we can interpret these excitations as
quasiparticles coupled to an effective staggered magnetic field $\tilde h_{\rm
  s}=\Delta\tilde\mu/g\mu_{\rm B}$, with $\tilde \mu$ playing the role of a
quasiparticle chemical potential. This equation gives the dispersion relation
for these single particle excitations, which can be regarded as constituting a
renormalized band, or bands as there are two branches. The term
magnetic polaron is sometimes used to describe these single particle
excitations in states of magnetic order, because of the analogy  with the
motion of a particle in a lattice to which it is strongly coupled, where the
excitation is termed a polaron.

   
The corresponding density of states of these free local quasiparticles on the
sublattice is 
\begin{equation} 
\tilde\rho_{0,\sigma}(\omega) \!=\!
\frac{1}{\sqrt{z_\uparrow z_\downarrow}}
\sqrt{\frac{\omega+\tilde\mu-\sigma\Delta\tilde\mu}
{\omega+\tilde\mu+\sigma\Delta\tilde\mu}}
\,\rho_0\!\left(\!\frac{\sqrt{(\omega+\tilde\mu)^2-\Delta\tilde\mu^2}}
{\sqrt{z_\uparrow  z_\downarrow}}\!\right),
\label{qpdos}
\end{equation} 
for $|\omega+\tilde\mu|> |\Delta\tilde \mu|$, and is zero otherwise. 
In the case of a half-filled band $\tilde\mu=0$ and there is a gap at the 
Fermi level $\epsF=0$.

To determine this local quasiparticle density of states in the presence of
the symmetry breaking staggered magnetic field we need to calculate
$z_\sigma$ and $\tilde\mu_{0,\sigma}$ for each spin type. Using the NRG
we can do this in two ways. As the DMFT-NRG calculations give us the self-energy
$\Sigma_{\sigma}(\omega)$ directly, we only need its value, and that of its first
derivative at $\omega=0$, to deduce both $z_\sigma$ and $\tilde\mu_{0,\sigma}$
using equation (\ref{rplat}). 
However, because the model is solved using an effective impurity model,
we can also deduce these quantities indirectly from the many-body energy
levels of the impurity on approaching the low energy  fixed
point \cite{HOM04}. This second method gives us not only a check on the
results of the direct method, but also allows to deduce some information about
the quasiparticle interactions, as we shall show in the next section.

\subsection{Calculation of Renormalized Parameters}

To describe how the renormalized parameters are deduced from the energy levels
of the NRG calculation, we need to outline how the NRG calculations are
carried out. Following the procedure introduced by Wilson \cite{KWW80a}, the
conduction band is logarithmically discretized and the model then converted
into the form of a one dimensional tight binding chain, coupled via an
effective hybridization $V_\sigma$ to the impurity 
at one end.  In this representation $K_\sigma(\omega)=|V_\sigma|^2
g^{(N)}_{0,\sigma}(\omega)$, where $g^{(N)}_{0,\sigma}(\omega)$ is the one-electron Green's
function for the first site of the isolated conduction electron chain of
length $N$.  The impurity Green's function for this discretized model then
takes the form, 
\begin{equation}
G^{\mathrm{imp}}_{\sigma}(\omega)={1\over
    \omega-\epsilon_{\mathrm{d}\sigma}-|
    V_\sigma|^2g^{(N)}_{0,\sigma}(\omega)-\Sigma_\sigma(\omega)}.
\label{rgfdmft}
\end{equation}

We can find the quasiparticle excitations of this model by expanding the self-energy
$\Sigma_\sigma(\omega)$ in the denominator of this equation 
to first order in $\omega$, and  write the result in the form,
\begin{equation}
G^{\mathrm{imp}}_{\sigma}(\omega)={z_\sigma\over
    \omega-\tilde\epsilon_{\mathrm{d}\sigma}-|\tilde
    V_\sigma|^2g^{(N)}_{0,\sigma}(\omega)+\ord{\omega^2}},
\label{rgfdmft}
\end{equation}
where 
\begin{equation}
\tilde\epsilon_{\mathrm{d}\sigma}=z_\sigma[\epsilon_{\mathrm{d}\sigma}+\Sigma_\sigma(0)],\quad
|\tilde V_\sigma|^2={z_\sigma}| V_\sigma|^2. 
\label{rp}
\end{equation}
We can then define a free quasiparticle  propagator, $\tilde
G_{0,\sigma}(\omega)$, viz
\begin{equation}
\tilde G^{\mathrm{imp}}_{0,\sigma}(\omega)={1\over
    \omega-\tilde\epsilon_{\mathrm{d}\sigma}-|\tilde V_\sigma|^2g^{(N)}_{0,\sigma}(\omega)},
\label{qpgfdmft}
\end{equation}
and interpret ${z_\sigma}$ as the local quasiparticle weight. 

In the NRG calculation the  many-body excitations are  calculated
iteratively, starting at the impurity site, and increasing the chain length
$N$ by one site with each iteration. When the matrices become too large to
handle, only the lowest 500-1500 states are kept at each iteration. The
many-body energy levels for the $N$th iteration and  the set of quantum
numbers $M$, $E_M(N)$, depend on the chain length $N$ and the discretization
parameter $\Lambda>1$. When $N$ becomes large these energy levels  go to zero
as $\Lambda^{-N/2}$.  We now conjecture that the lowest single particle
$E^{\sigma}_{p}(N)$ and single hole excitations $E^{\sigma}_{h}(N)$ determined from the NRG
many-body excitations correspond to quasiparticle excitations. If this is the
case then they should correspond to the poles of the quasiparticle Green's
function given in equation (\ref{qpgfdmft}), with values of $\tilde V_\sigma$
and $\tilde\epsilon_{\mathrm{d}\sigma}$, which are independent of $N$ as
$N\to \infty$.  We can test this hypothesis by substituting the values,
$\omega=E^{\sigma}_{p}(N)$ and  $\omega=E^{\sigma}_{h}(N)$, into the equation, 
 \begin{equation}
    \omega-\tilde\epsilon_{\mathrm{d}\sigma}-|\tilde
    V_\sigma|^2g^{(N)}_{0,\sigma}(\omega)=0 ,
\label{}
\end{equation}  
and deduce values of $\tilde V_\sigma$ and
$\tilde\epsilon_{\mathrm{d}\sigma}$, which will in general depend upon
$N$. From these we can deduce $z_\sigma=|\tilde V_\sigma /V_\sigma|^2$ and 
$\tilde\mu_{0,\sigma}=-\tilde\epsilon_{\mathrm{d}\sigma}$, which will also
depend upon $N$, but if the lowest single particle excitations of the system
do correspond to free quasiparticles, the values of $z_\sigma$ and
$\tilde\mu_{0,\sigma}$ will become independent of $N$ for large $N$. It should
be noted that we need both the particle and hole excitations for each spin to
determine the four renormalized parameters. The parameters corresponding to spin
up involve the particle excitations with spin up and the hole excitations with spin down.

\begin{figure}[!htbp]
\centering
\includegraphics[width=0.45\textwidth]{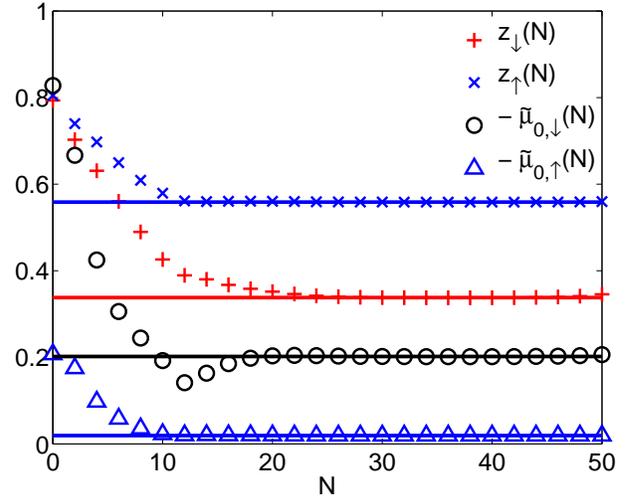}
\caption{(Color online) The $N$-dependence of the renormalized parameters $z_\sigma$ and
  $\tilde\mu_{0,\sigma}$ for $U=3$ and $x=0.9$.} 
\label{renparNRG}
\end{figure}

That parameters can be found, which are independent of $N$ for large $N$, can
be seen in figure \ref{renparNRG}, where we take the results of a 
$K_\sigma(\omega)$ and $\mu_{\sigma}$ from the antiferromagnetic
self-consistent solution for the Hubbard model with $U=3$ and 10\% doping,
using a value for the discretization parameter $\Lambda=1.8$. 

\noindent
It can be seen that after about 25 iterations all the values deduced for
$z_\sigma$ and $\tilde \mu_{0,\sigma}$ become independent of $N$. In the next
section, where we compare these results with the corresponding values deduced
directly from the self-energy, we get further confirmation that the values deduced really
do describe the quasiparticle excitations of the lattice model. 

When two or more quasiparticles are excited from the interacting ground state,
there will be an interaction between them. For the Anderson  impurity model
this interaction will be local and can be expressed as $\tilde U$, a
renormalized
value of the original interaction of the `bare' particles. The value of
$\tilde U$ can be deduced by looking at lowest lying two-particle excitations derived
from NRG calculation. These could either be two-particle excitations,
$E^{\uparrow,\downarrow}_{pp}(N)$, two-hole excitations,
$E^{\downarrow,\uparrow}_{hh}(N)$ or a particle-hole excitation
$E^{\uparrow,\uparrow}_{ph}(N)$. By looking at the difference between a
two-particle excitation and two single particle excitations,
$E^{\uparrow,\downarrow}_{pp}(N)-E^{\uparrow}_{p}(N)-E^{\downarrow}_{p}(N)$,
as a function of $N$ we can deduce an effective interaction $\tilde
U^{\uparrow,\downarrow}_{pp}(N)$ between these two quasiparticles, as has been
described fully earlier for the standard Anderson model \cite{HOM04}. In a similar way we
can deduce an effective interaction between two holes,  $\tilde
U^{\downarrow,\uparrow}_{hh}(N)$, or a particle and hole, $-\tilde
U^{\uparrow,\uparrow}_{ph}(N)$. To be able to define a single quasiparticle
interaction $\tilde U$, not only 
must $\tilde U^{\uparrow,\downarrow}_{pp}(N)$, $\tilde
U^{\downarrow,\uparrow}_{hh}(N)$ and $\tilde U^{\uparrow,\uparrow}_{ph}(N)$,
give values which are independent of $N$ for large $N$,  these values must be
equal so $\tilde U^{\uparrow,\downarrow}_{pp}=\tilde
U^{\downarrow,\uparrow}_{hh}=\tilde U^{\uparrow,\uparrow}_{ph}=\tilde U$.

\begin{figure}[!htbp]
\centering
\includegraphics[width=0.45\textwidth]{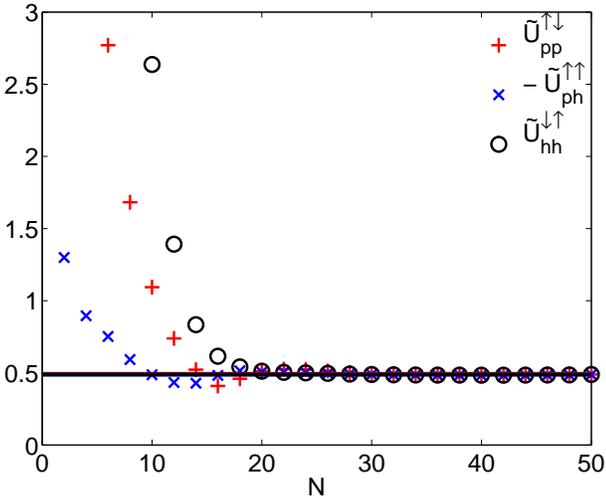}
\caption{(Color online) The $N$-dependence of the renormalized particle-particle,
particle-hole and hole-hole interactions for $U=6$ and $x=0.9$,
showing that they converge to a unique value $\tilde U$.}
\label{renparNRGtU}
\end{figure}
\noindent
In figure \ref{renparNRGtU} we give the values of $\tilde
U^{\uparrow,\downarrow}_{pp}(N)$,  $\tilde U^{\downarrow,\uparrow}_{hh}(N)$
and $\tilde U^{\uparrow,\uparrow}_{ph}(N)$ as deduced from DMFT-NRG
calculation  for the Hubbard model in  an antiferromagnetic state  with $U=6$,
10\% doping and  $\Lambda=1.8$. It can be seen
that the three sets of results settle down to a common value $\tilde U$.

We can go further and identify $\tilde U$ with the local quasiparticle  4-vertex interaction
for the effective impurity model,
\begin{equation}
\tilde U=z_\uparrow z_\downarrow\Gamma_{\uparrow,\downarrow,\downarrow,\uparrow}(0,0,0,0),
\end{equation}
where   $\Gamma_{\uparrow,\downarrow,\downarrow,\uparrow}(\omega_1,
\omega_2,\omega_3,\omega_4)$ is the total 4-vertex at the impurity site, which
is equal to the same quantity for a site in the lattice model.
With this interpretation it is possible to identify  these  parameters
 with those used in a renormalized perturbation
expansion. The parameters, $V$, $\epsilon_{{\rm d},\sigma}$ and
$U$, together with $g_{0,\sigma}^{N}(\omega)$, specify the effective impurity
 model.  
 The  renormalized parameters, $\tilde V$, $\tilde\epsilon_{{\rm d},\sigma}$ and
$\tilde U$, together with $g_{0,\sigma}^{N}(\omega)$, can be used as an alternative way
of specifying this model. The renormalized perturbation theory (RPT) is set up
by expanding the self-energy to order $\omega$, as earlier,  but retaining
all the higher order correction terms in a remainder term, 
 \begin{equation}
\Sigma_{\sigma}(\omega)=\Sigma_{\sigma}(0)+\omega\Sigma'_{\sigma}(0)+\Sigma^{\rm
  rem}_{\sigma}(\omega),
\label{rem} 
\end{equation}
where $\Sigma^{\rm rem}_{\sigma}(\omega)$ is the remainder term. On substituting
this into the equation for the impurity Green's function in equation
(\ref{gfdmft}), we can deduce a general expression for the quasiparticle
Green's function in the form,
\begin{equation}
\tilde G^{\mathrm{imp}}_{\sigma}(\omega)=\frac{1}
{\omega-\tilde\epsilon_{\mathrm{d}\sigma}-\tilde K_\sigma(\omega)-\tilde\Sigma_\sigma(\omega)},
\label{qpgf}
\end{equation}
where $\tilde K_\sigma(\omega)=z_\sigma K_\sigma(\omega)$ and
$\tilde\Sigma_{\sigma}(\omega)=z_\sigma\Sigma^{\rm rem}_{\sigma}(\omega)$
plays the 
role of a renormalized self-energy. A diagrammatic perturbation theory  can
then be carried out for $\tilde\Sigma_{\sigma}(\omega)$ in terms of the free
quasiparticle propagators, 
with additional diagrams arising from counter terms, which are required 
to prevent over-counting (renormalization conditions) \cite{ryder,Hew93,Hew01}. 
This form of perturbation theory is valid for all
energy scales but is particularly effective for calculating the low energy
terms arising from the quasiparticle interactions. For the symmetric Anderson
impurity model it has been shown that this perturbation theory taken to second order in
$\tilde U$, gives the exact spin and charge susceptibilities at $T=0$, and
the exact $T^2$ contribution to the conductivity \cite{Hew93}.

Because, within DMFT, the self-energy for the lattice is the same as that for
the effective impurity, we can equally well use the effective impurity model
to calculate it. This means that we can use the renormalized perturbation
theory for the effective impurity model to estimate the correction terms to
the free quasiparticle picture arising from the quasiparticle
interactions.

\subsection{Local Quasiparticle Weight}

\begin{figure}[!htbp]
\begin{center}
\includegraphics[width=0.45\textwidth]{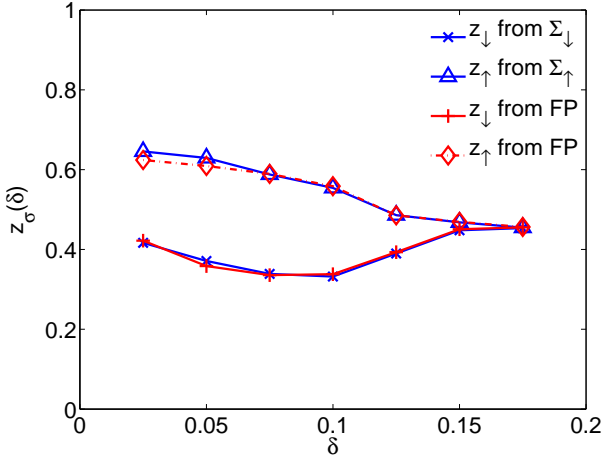}
\end{center} 
\vspace*{-0.5cm}
\caption{(Color online) The local quasiparticle weight $z_{\sigma}$ as deduced
  directly from the self-energy and also from the impurity fixed point (FP)
  for $U=3$ and various dopings.} 
\label{qpweightU3}
\end{figure}

We now consider the values of the local quasiparticle weight factor $z_{\sigma}$,
commonly known also as the wavefunction renormalization factor. This is an
important factor in determining the parameters needed to describe the low
energy behavior of the system.  When there is no ${\vk}$-dependence of the
self-energy, as is the case for infinite dimensional models and DMFT, the
effective mass of the quasiparticles in the paramagnetic state is proportional
to $1/z_\sigma$. We show later that in the antiferromagnetic state 
the expression is more complicated and depends both on $z_{\sigma}$ and the
renormalized chemical potential $\tilde\mu_{0,\sigma}$.
We have determined $z_\sigma$ from the NRG results by the two methods
described and give the values of $z_\sigma$ deduced  for both spin types as a
function of doping  in figure \ref{qpweightU3}.  
The results are for the case $U=3$, where there is antiferromagnetic order and
the external staggered field has been set to zero. 
\noindent
It can be seen that there is a reasonable agreement between the values obtained by the two
different methods of calculation. Visible differences can be attributed to the
inaccuracies when numerically computing the derivative of the self-energy,
whose calculation involves a broadening procedure.
When the system is doped but still ordered we have $z_{\uparrow}\neq
z_{\downarrow}$, and the renormalization effects are stronger 
for the minority (down) spin particles on the sublattice. This is similar to
the results we found for a doped  Hubbard model in a paramagnetic state in the
presence of a strong 
uniform  magnetic field \cite{BH07bpre}. For a certain range of dopings the values of
$z_{\uparrow}$ and $z_{\downarrow}$ do not vary much.  The tendency is that
$z_{\downarrow}$ first decreases and later increases, whereas $z_{\uparrow}$
decreases over the whole range until both of them merge at the
doping point where the antiferromagnetic order disappears. 

\begin{figure}[!htbp]
\begin{center}
\includegraphics[width=0.45\textwidth]{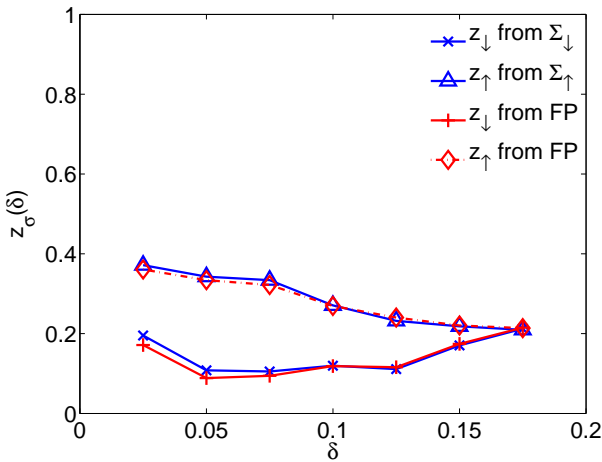}
\end{center} 
\vspace*{-0.5cm}
\caption{(Color online) The local quasiparticle weight $z_{\sigma}$ as deduced
  directly from the self-energy and from the impurity fixed point (FP) for
  $U=6$ and various dopings.} 
\label{qpweightU6}
\end{figure}

The results for the corresponding case with $U=6$, a value which is larger than the
bandwidth, are shown in figure \ref{qpweightU6}. 

\noindent
On the whole the behavior is quite similar to that for the case $U=3$,
only that the renormalization  effects are more pronounced.
For a range of dopings the local quasiparticle weights do
not change much and have the same tendency as described above. The
implications for the spectral quasiparticle weight and the effective mass
enhancement will be discussed in detail later. 



\subsection{Renormalized chemical potential}

\begin{figure}[!htbp]
\begin{center}
\includegraphics[width=0.45\textwidth]{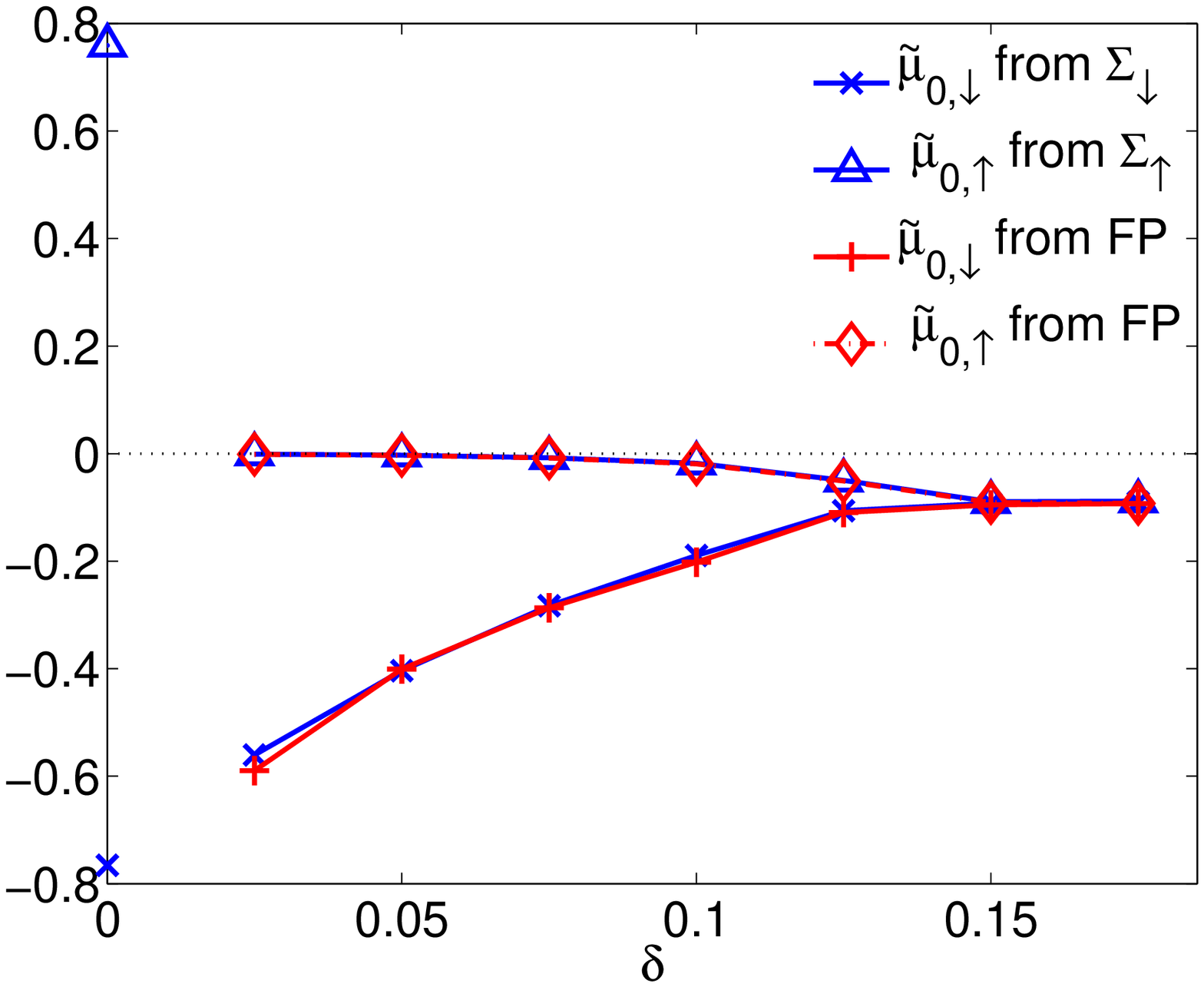}
\includegraphics[width=0.45\textwidth]{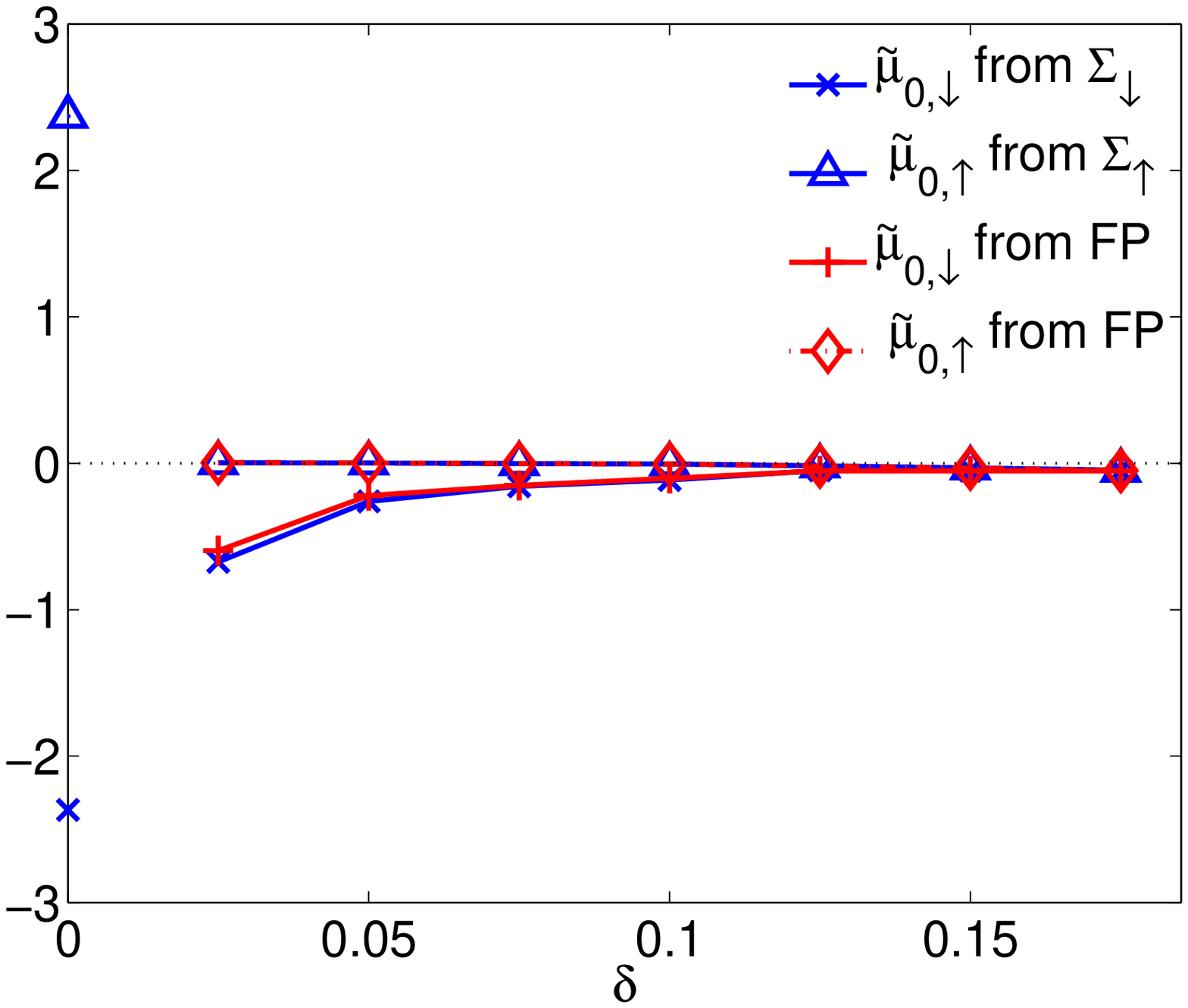}
\end{center} 
\vspace*{-0.5cm}
\caption{(Color online) The renormalized chemical potential
  $\tilde\mu_{0,\sigma}$ as deduced directly from the self-energy and  from
 the impurity fixed point (FP) for various dopings for $U=3$ (upper panel) and
 $U=6$ (lower panel).} 
\label{tmuU3}
\end{figure}

In figure \ref{tmuU3}  we give the  results for the renormalized
chemical potential, $\tilde\mu_{0,\sigma}$ [defined in equation (\ref{rplat})
and (\ref{rp})], for the two spin types in the spontaneously ordered
antiferromagnetic states for $U=3$ and $U=6$ for a range of dopings. 

\noindent
The values calculated by the two different methods can be seen to be in good
agreement here, as well. We have added the values for the half filled case.
These were calculated from the self-energy in the gap at $\omega=0$. The
general behavior  
of the values for $\tilde\mu_{0,\sigma}$ for the case with $U=6$ is very
similar to the case with smaller $U$
 
The renormalized chemical potential $\tilde\mu_{0,\sigma}$  is an important
parameter in specifying the form of the local sublattice quasiparticle spectral
density $\tilde\rho^{0}_{\sigma}(\omega)$. From equation (\ref{qpdos}) it can
be seen that, as $\omega\to-\tilde\mu_{0,\sigma}$,
$\tilde\rho_{0,\sigma}(\omega)$ behaves  asymptotically  as 
\begin{equation}
  \tilde\rho_{0,\sigma}(\omega)\sim\frac1{\sqrt{\omega+\tilde\mu_{0,\sigma}}},
\end{equation}
so the quasiparticle density of states has a square root singularity at
$\omega=-\tilde\mu_{0,\sigma}$. On the other hand, however, as
$\omega\to-\tilde\mu_{0,-\sigma}$, $\tilde\rho_{0,\sigma}(\omega)$ behaves
as 
\begin{equation}
  \tilde\rho_{0,\sigma}(\omega)\sim {\sqrt{\omega+\tilde\mu_{0,-\sigma}}},
\end{equation}
so the quasiparticle density of states goes to zero at
$\omega=-\tilde\mu_{0,-\sigma}$. Between the two points,
$\omega=-\tilde\mu_{0,\sigma}$ and $\omega=-\tilde\mu_{0,-\sigma}$, the
quasiparticle density of states has a gap of magnitude $2\Delta\tilde \mu$.  
As can be seen in figure \ref{tmuU3} this free quasiparticle gap decreases
with increasing doping and closes in the paramagnetic state. If we take into account
the values at half filling we see a strong reduction of $2\Delta\tilde \mu$,
when doping the system. We also see that $\tilde\mu_{0,\uparrow}$ drops to
small negative values for finite hole doping, which corresponds to the fact that
the Fermi level then lies within the lower band.
These features will be seen clearly in the figures  presented in the next
section, where we compare the quasiparticle densities of states with the full
local  spectral densities calculated  from the DMFT-NRG.  

\subsection{The Quasiparticle Interaction}
\begin{figure}[!htbp]
\begin{center}
\includegraphics[width=0.45\textwidth]{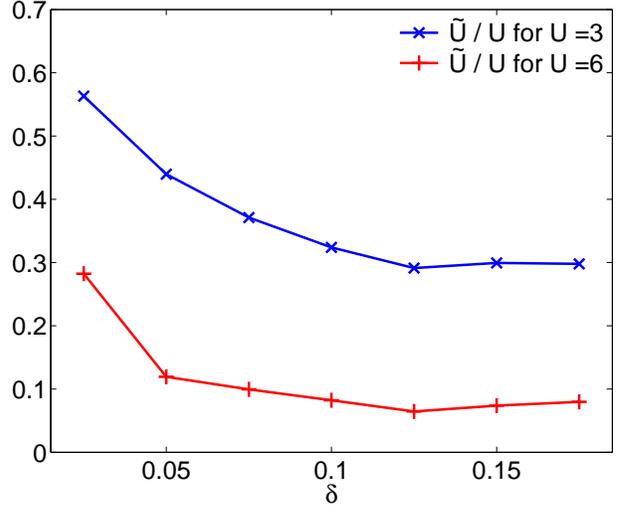}
\end{center} 
\vspace*{-0.5cm}
\caption{(Color online) The renormalised quasiparticle interaction $\tilde
  U/U$  as deduced from the impurity fixed point for various dopings and $U=3,6$.}  
\label{tU_U3}
\end{figure}

The quasiparticles can be further characterized by an effective interaction
$\tilde U$ as described before. In figure \ref{tU_U3} we plot the doping
dependence of the renormalized interaction over a range of dopings and $U=3$
and $U=6$.

\noindent
We can see that in both cases the values decrease with increasing
doping. Hence, the effective quasiparticle  
interaction is stronger for a smaller hole density. For a certain range of
dopings, however, $\tilde U$ does not vary much. We can also see that the
ratio $\tilde U/U$ for the effective interaction assume smaller values the
larger the bare $U$ becomes.  Also the absolute value of $\tilde U$,
i.e. without the scaling with $U$ as in figure \ref{tU_U3},  is smaller for
larger bare $U$ for the full range of dopings. 
This effect of smaller quasiparticle interactions for the stronger coupling
case can be seen as sharper quasiparticle peaks for larger $U$ as will be
discussed in the next section.  

\section{Spectra and Quasiparticle Bands}
\subsection{Local Spectra}
\begin{figure}[!htbp]
\begin{center}
\includegraphics[width=0.45\textwidth]{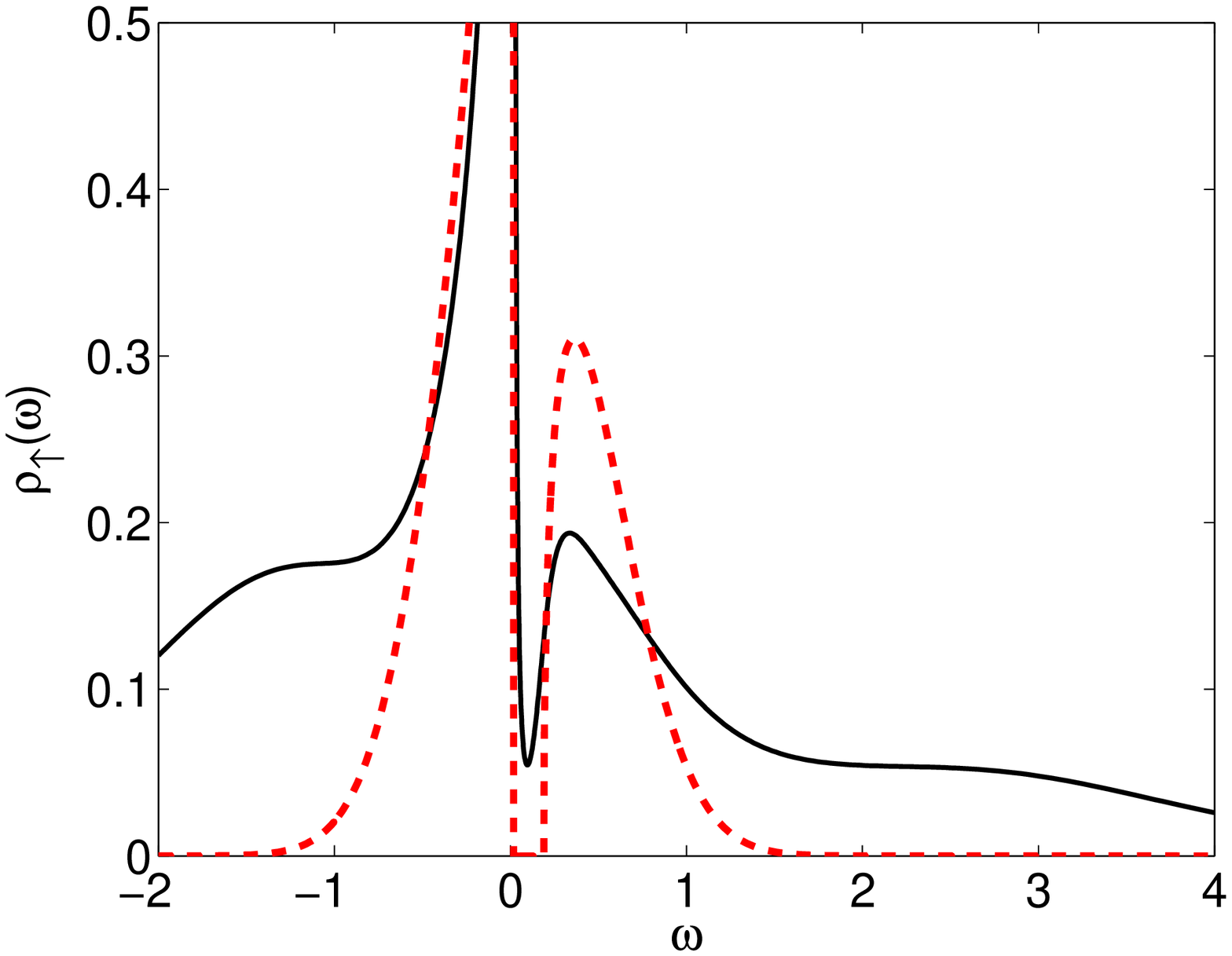}
\includegraphics[width=0.45\textwidth]{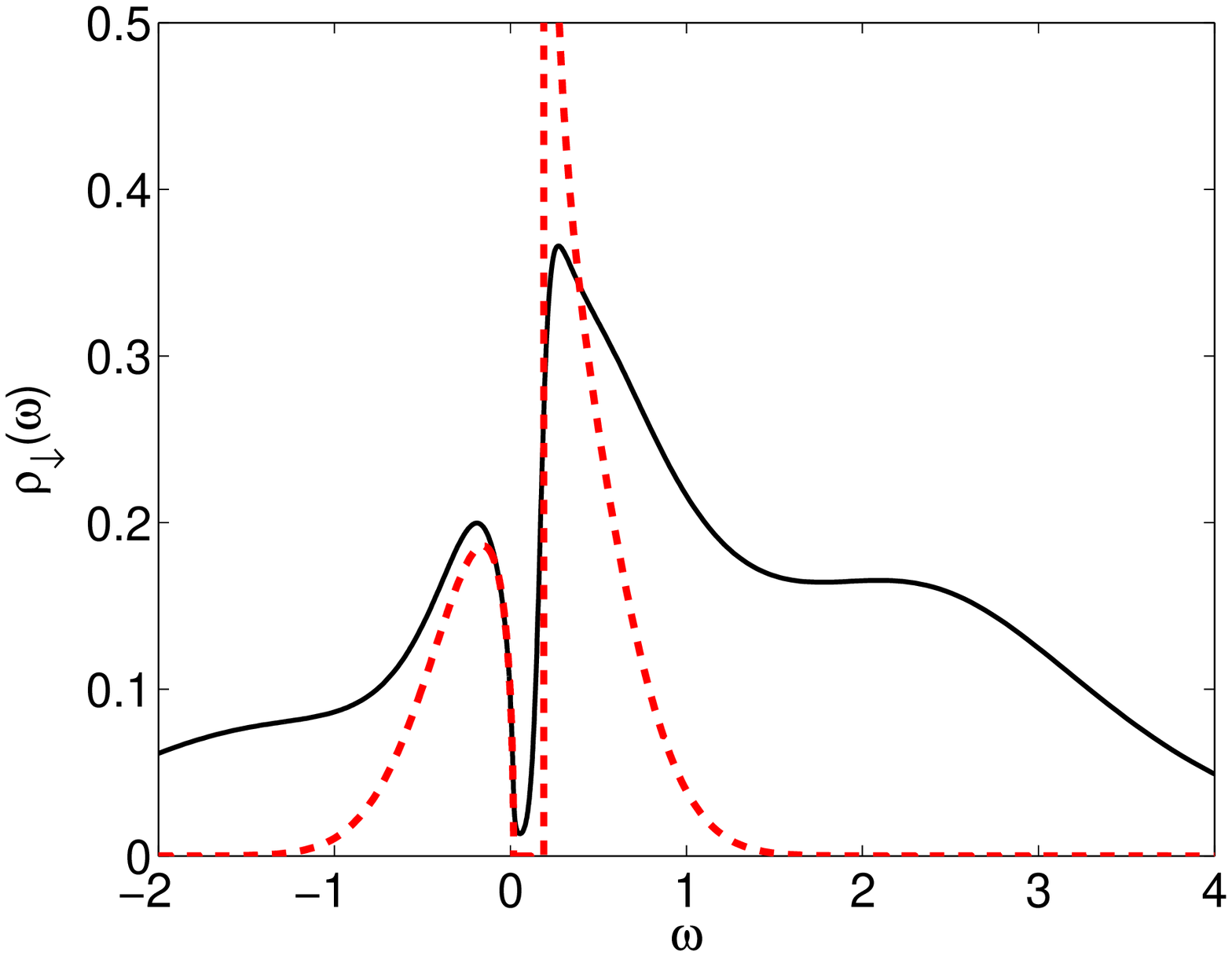}
\end{center} 
\vspace*{-0.5cm}
\caption{(Color online) The free local quasiparticle spectrum (dashed line) in
  comparison with DMFT-NRG spectrum for $x=0.9$ and $U=3$ for the spin-up
  electrons (upper panel) and spin-down electrons (lower panel).}
\label{qpspecU3x0.9}
\end{figure}

In this section we examine how well the local sublattice quasiparticle density of
states $\tilde\rho_{0,\sigma}(\omega)$, 
evaluated from  equation (\ref{qpdos}) with the renormalised parameters,
describes the low energy features seen in the  local spectral density 
$\rho_{\sigma}(\omega)$ calculated from the DMFT-NRG. At half filling there is
a gap at the Fermi level, so there are no single particle excitations 
in the immediate neighbourhood of the Fermi level, and this is not a very 
interesting case to consider. We look in detail at the case of $10\%$ doping
where the Fermi level lies at the top of the lower band, and consider
the two cases $U=3$ and $U=6$. In the upper panel of figure \ref{qpspecU3x0.9} 
we compare the spectral density $\rho_{\uparrow}(\omega)$ with the
corresponding quantity $z_\uparrow\tilde\rho_{0,\uparrow}(\omega)$, from the
quasiparticle density of states. 

\begin{figure}[!htbp]
\begin{center}
\psfrag{w}{$\omega$}
\psfrag{mw}{$-\omega$}
\psfrag{eps1}{$\epsilon_1$}
\psfrag{eps2}{$\epsilon_2$}
\psfrag{wmmu}{$-\omega+|\tilde\mu_{0,\uparrow}|$}
\psfrag{mu1}{$|\tilde\mu_{0,\uparrow}|$}
\psfrag{mu2}{$|\tilde\mu_{0,\downarrow}|$}
\includegraphics[width=0.45\textwidth]{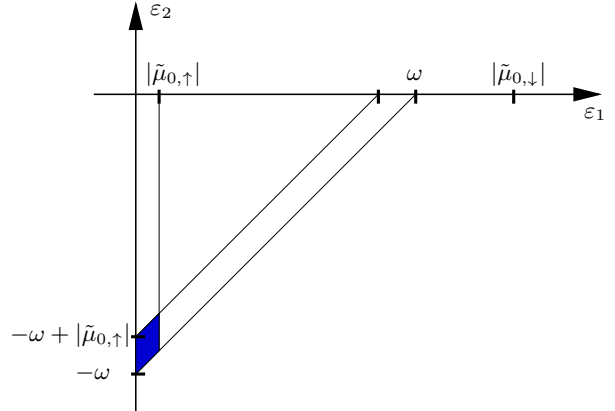}
\end{center} 
\vspace*{-0.5cm}
\caption{(Color online) Integration region in the $(\epsilon_1,\epsilon_2)$-plane for the
  imaginary part of the self-energy. The original triangle region
  $(0,\omega,-\omega)$ for integration in equation (\ref{afmrensig2})
  is reduced in the gap region,
  $|\tilde\mu_{0,\uparrow}|<\omega<|\tilde\mu_{0,\downarrow}|$, to the small shaded
  region shown in the figure.}
\label{intregion}
\end{figure}

\noindent
We see that the behavior near the Fermi level ($\omega=0$), and the singular feature
seen in the lower branch of $\rho_{\uparrow}(\omega)$, are well reproduced by
the quasiparticle density of states.  Above the Fermi level there is a peak in
the quasiparticle density of states  similar to that in the full spectrum but
somewhat more pronounced. Above the Fermi level and below the upper peak there
is a pseudo-gap region. In the free quasiparticle spectrum it is a definite
gap. In the spectrum  calculated from the direct NRG evaluation it appears as
a pseudo-gap, with rather small spectral weight just above the Fermi
level. From the direct DMFT-NRG calculations, due to the broadening features introduced to
obtain a continuous spectrum, it is not always possible  to say definitively
whether there is a true gap above the Fermi level or not.  To resolve this
question we can appeal to the renormalised perturbation theory to look at the
corrections to the quasiparticle density of states arising from the
quasiparticle interactions. A calculation of the imaginary
part of the renormalised self-energy $\tilde\Sigma_\sigma(\omega)$ 
to order $\tilde U^2$ should be sufficient to settle this issue.
The imaginary part of the second order diagram for the renormalised
self-energy in the limit $T\to 0$ for $\omega>0$ is given by  
\begin{eqnarray}
 \Imag\tilde\Sigma^{(2)}_{\sigma}(\omega)&=&\pi\tilde
  U^2\integral{\epsilon_1}0{\omega}\!\!
 \integral{\epsilon_2}0{-\omega}\tilde\rho_{0,\sigma}(\epsilon_1)
 \tilde\rho_{0,-\sigma}(\omega-\epsilon_1+\epsilon_2)  \nonumber  \\
&&\times \;\; \tilde\rho_{0,-\sigma}(\epsilon_2)
\theta(\omega-\epsilon_1+\epsilon_2),
 \label{afmrensig2} 
\end{eqnarray}
 where $\tilde\rho_{0,\sigma}(\epsilon)$ is the free quasiparticle density of
states. The integration area is a triangle in the
$(\epsilon_1,\epsilon_2)$-plane as shown in figure \ref{intregion}. 

\noindent
To analyze the behavior of $\Imag\tilde\Sigma^{(2)}_{\sigma}(\omega)$ in the
regime $|\tilde\mu_{0,\uparrow}|<\omega<|\tilde\mu_{0,\downarrow}|$ we have to
study where the integrand is non-zero taking into account that
$\tilde\rho_{0,\sigma}(\epsilon)= 0$  for
$|\tilde\mu_{0,\uparrow}|<\epsilon<|\tilde\mu_{0,\downarrow}|$. The only
non-zero contribution comes from the small shaded region in figure 
 \ref{intregion}, which leads to the estimate,
\begin{equation}
  \Imag\tilde\Sigma^{(2)}_{\sigma}(\omega)\simeq \pi\tilde
  U^2\tilde\rho_{0,\sigma}(0)
\tilde\rho_{0,-\sigma}(-\omega)\tilde\rho_{0,-\sigma}(0) \tilde\mu_{0,\uparrow}^2.
\end{equation}
When $\tilde\mu_{0,\uparrow}$ is small, which occurs when the lower edge of
the gap in the quasiparticle density of states is very  near  the Fermi level,
this contribution to the imaginary part of the renormalized self-energy will
be finite but small.  It decreases with $\omega$ due the behavior of
$\tilde\rho_{0,-\sigma}(-\omega)$. 
Based on this argument we conclude that  there is a small, but finite imaginary
part of the self-energy in the free quasiparticle gap $2\Delta\tilde\mu$,
when it lies above the Fermi level, giving rise to a finite spectral weight
there.  However,  this spectral weight is very small close to the lower edge
of the free quasiparticle density of states, when this edge  lies only just
above the Fermi level.

In the lower panel of figure \ref{qpspecU3x0.9} we compare the
$\rho_{\downarrow}(\omega)$ with
$z_\downarrow\tilde\rho_{0,\downarrow}(\omega)$. We see that in this case also 
the quasiparticle density of states reproduces well the
spectrum in the region of the Fermi level and the peak structure in the lower
band, which is non-singular in this case. The position of the peak above the
Fermi level is also well reproduced, but the peak  in the free quasiparticle
density
of states is singular, whereas that in  the DMFT-NRG results is not.
  We would expect to lose this singularity in the free quasiparticle density
  of states once the quasiparticle
scattering is taken into account and the renormalized self-energy is included.
It is possible also that the peak above the Fermi level in the DMFT-NRG
spectrum should be  sharper, as there is some tendency for the broadening
introduced in this approach to flatten peaked features in regions away from the
Fermi level.  
The spectral weight in the pseudo-gap is even smaller than in the case for the
spin-up electrons, particularly in the region of the gap that lies  closest to
the Fermi level. This is qualitatively in line with the conclusions based on
the renormalized perturbation theory estimate of the effects of the
quasiparticle scattering.

\begin{figure}[!htbp]
\begin{center}
\includegraphics[width=0.45\textwidth]{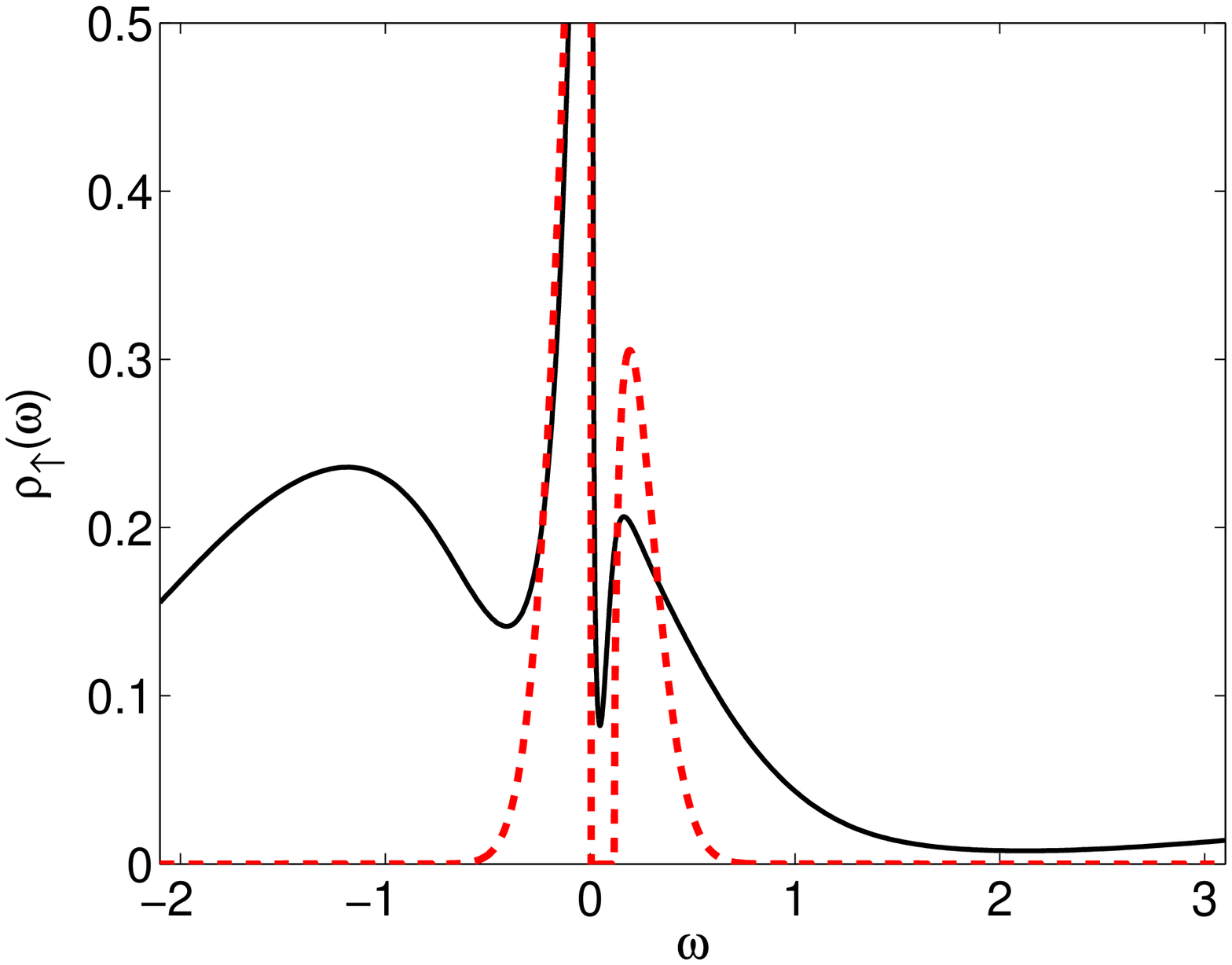}
\hspace{1cm}
\includegraphics[width=0.45\textwidth]{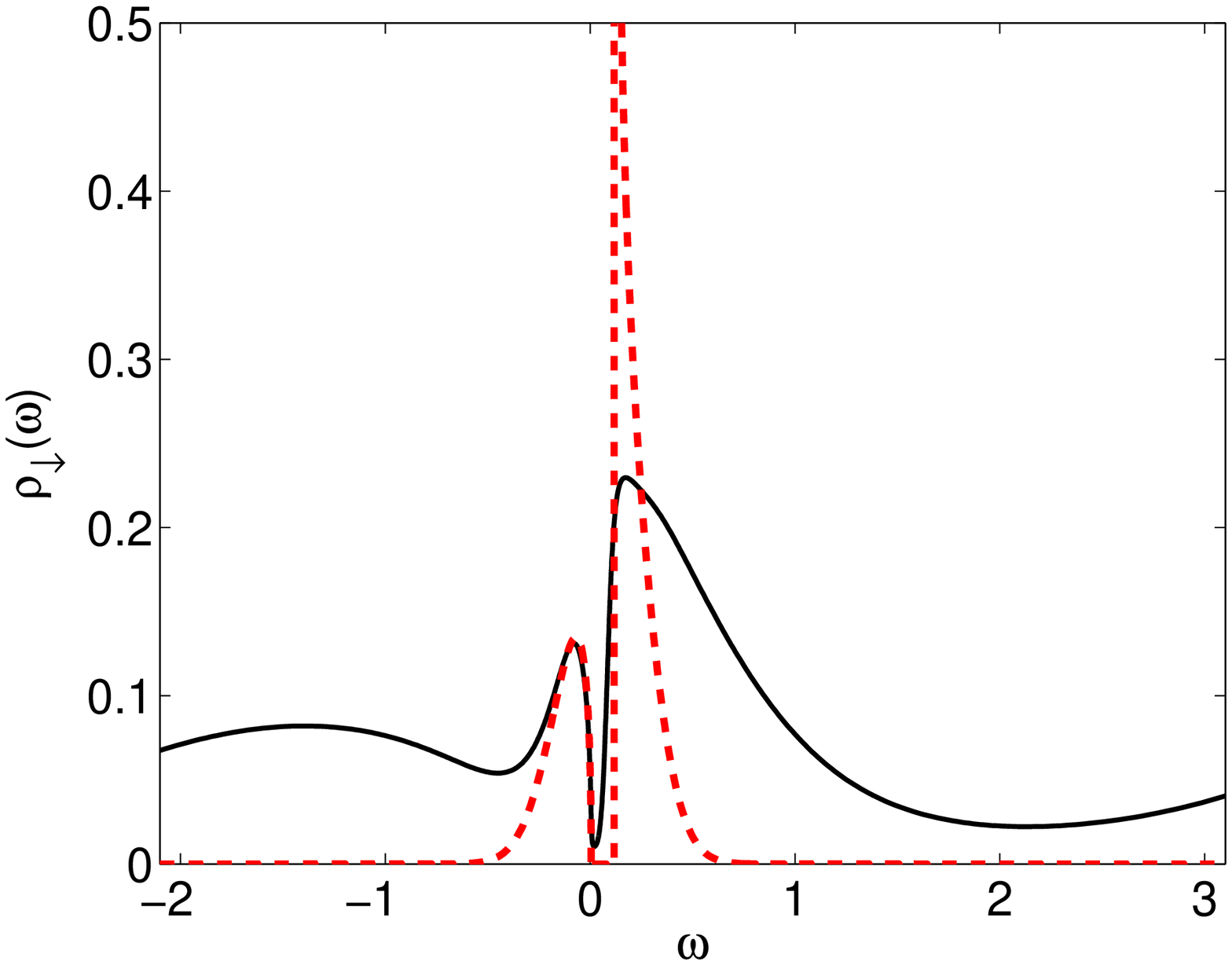}
\end{center} 
\vspace*{-0.5cm}
\caption{(Color online) The free quasiparticle spectrum (dashed line) in comparison with
  DMFT-NRG spectrum for $x=0.9$ and $U=6$ for the spin-up electrons
(upper panel) and spin-down electrons (lower panel).}
\label{qpspecU6x0.9}
\end{figure}

We see very similar features in the spectra for the case $U=6$ and also 10\%
doping shown in figure \ref{qpspecU6x0.9}. 

\noindent
Here, the peaks near the Fermi level are a bit sharper. The observations
made on the comparison of the quasiparticle and DMFT-NRG spectra apply equally
well to this case. In addition to the low energy features charge peaks
corresponding to the Hubbard bands appear. The lower one can be identified in 
the full spectra, whereas the upper Hubbard peak is not seen on the energy
scale shown. The quasiparticle density of states does not contain
information about these features at higher energy. 

\subsection{$\vk$-resolved  Spectra}
We can learn more  about the low energy single particle excitations by looking at
the spectral density of the Green's function $\underline{G}_{\vk,\sigma}(\omega)$
in equation (\ref{kgf}) for a given wave-vector ${\vk}$. With the
self-energies $\Sigma_{\sigma}(\omega)$  calculated within the DMFT-NRG
approach all elements of this matrix can be evaluated. The local spectra and
self-energies are spin-dependent in the doped broken symmetry state, however,
the free quasiparticle bands  $E^0_{\vk,\pm}$ [equation (\ref{qpex})] do not
depend on the spin. Here, we focus on the diagonal part of
$\underline{G}_{\vk,\sigma}(\omega)$ corresponding to the $A$ sublattice, 
\begin{equation}
  G_{\vk,\sigma}(\omega)=\frac{\zeta_{-\sigma}(\omega)}
{\zeta_{\sigma}(\omega)\zeta_{-\sigma}(\omega)-\epsilon_{\vk}^2}.
\label{nkgf}
\end{equation}
The weights of the quasiparticle excitations in this case
depend on the spin corresponding to the sublattice properties.
We note that one can also analyze the quasiparticle bands differently, for
instance, from the $\vk$-resolved spectra and the diagonal form of
$\underline{G}_{\vk,\sigma}(\omega)$. The form of the quasiparticle bands
remains unchanged then, but the weights differ and do not depend on the spin
$\sigma$ in that case. 

We first of all look at the
Fermi surface which is  the locus of
the ${\vk}$-points at the Fermi level ($\omega=0$) where the Green's function has poles.
The conduction electron  energy $\epsilon_{{\vkF}}$ at these point is given by 
 \begin{equation}
 \epsilon^2_{{\vkF}}=(\mu_\uparrow-\Sigma_\uparrow(0))(\mu_\downarrow-\Sigma_\downarrow(0)).
\label{ifs} 
\end{equation}
By Luttinger's theorem, the volume of the Fermi surface for the interacting
system must equal  that for 
the non-interacting system with the same density. As the self-energy depends
only on $\omega$, the two Fermi surfaces must also have the same shape, and
therefore  must be identical.
The Fermi surface of the non-interacting system is given by  $\epsilon_{
{\vkF}}=\mu_0$, where $\mu_0$ is  the chemical potential of the
non-interacting system in the absence of any applied field for the given
density.  For this to be identical with that given in equation 
(\ref{ifs}),
 \begin{equation}
(\mu_\uparrow-\Sigma_\uparrow(0))(\mu_\downarrow-\Sigma_\downarrow(0)) =\mu_0^2.
\end{equation}
We can check that this relation indeed holds from  our results for
$\Sigma_\sigma(\omega)$ and $\mu_\sigma$, independent of the value of $U$,
or in the case of an applied staggered field, independent of the field value.
This relation implies that the total number of electrons per site $n$ can be
calculated from an integral over the non-interacting density of states,
 \begin{equation}
n=2\int \limits_{-\infty}^{\mu_0}\rho_0(\omega)d\omega,
\label{ntotal}
\end{equation}
where in the hole doped case 
$\mu_0=-\sqrt{\bar\mu_\uparrow\bar\mu_\downarrow}$ and $\bar\mu_\sigma= 
\mu_\sigma-\Sigma_\sigma(0)$.

To relate this result to the quasiparticle picture, we expand the self-energy
in equation (\ref{nkgf}) to first order in $\omega$, 
but retain the remainder term, $\Sigma^{R}_\sigma(\omega)$ as in equation
(\ref{rem}).
The  Green's function can be rewritten in the form, 
\begin{equation}
  \tilde G_{\vk,\sigma}(\omega)=\frac{\tilde\zeta_{-\sigma}(\omega)}
{\tilde\zeta_{\sigma}(\omega)\tilde\zeta_{-\sigma}(\omega)-\tilde\epsilon_{\vk}^2},
\label{rengfct}
\end{equation}
where
$\tilde\zeta_{\sigma}(\omega)=\omega+\tilde\mu_{0,\sigma}-\tilde\Sigma_{\sigma}(\omega)$.
We define a quasiparticle Green's function  $\tilde G_{\vk,\sigma}(\omega)$
via   $z_\sigma \tilde G_{\vk,\sigma}(\omega)=G_{\vk,\sigma}(\omega)$. The
renormalized self-energy vanishes,  $\tilde\Sigma_{\sigma}(\omega)=0$, for the
free quasiparticle Green's function  $\tilde G^{(0)}_{\vk,\sigma}(\omega)$,
which can be separated into two independent branches  of  free quasiparticles, 
\begin{equation}
  \tilde G^{(0)}_{\vk,\sigma}(\omega)=
\frac{ u_{+}^{\sigma}(\epsilon_{\vk})}{\omega-E^0_{\vk,+}}
+\frac{u_{-}^{\sigma}(\epsilon_{\vk})}{\omega-E^0_{\vk,-}},
\end{equation}
where $E^0_{\vk,\pm}$ was defined in equation (\ref{qpex}) and the weights are
given by
\begin{equation}
u_{\pm}^{\sigma}(\epsilon_{\vk})=\frac12\left(1\mp\sigma\frac{\Delta\tilde\mu}
{\sqrt{\Delta\tilde\mu^2+\tilde\epsilon_{\vk}^2}}\right).
\label{weights}
\end{equation} 
This is similar in form to mean field theory, which would correspond to
putting $z_{\sigma}=1$, and $\Delta\tilde\mu=Um_{\rm mf}$, where $m_{\rm
mf}$ is the mean field sublattice magnetization. The spin dependent
contribution in (\ref{weights})  which arises from the second term is most
marked in the region near the Fermi level. It should be noted that the
quasiparticle excitations $E^0_{\vk,\pm}$ and weights
$u_{\pm}^{\sigma}(\epsilon_{\vk})$ here are defined by expanding the
self-energy at $\omega=0$. This is so that they correspond to the
free quasiparticles in the renormalized perturbation theory which have an
infinite lifetime. 

The spectral density $\tilde\rho^{(0)}_{\vk}(\omega)$ for this free
quasiparticle Green's function is a set of delta-functions,
\begin{equation}
 \tilde\rho^{(0)}_{{\vk},\sigma}(\omega)=
 u_{+}^{\sigma}(\epsilon_{\vk})\delta(\omega-E^0_{\vk,+})
+ u_{-}^{\sigma}(\epsilon_{\vk})\delta(\omega-E^0_{\vk,-} ).
\label{qpdeltaexc}
\end{equation}
On the Fermi surface $E^0_{\vk,-}=0 $, which is consistent with the result
for the Fermi surface given in equation (\ref{ifs}). Summing over ${\vk}$ gives the
local quasiparticle density of states in equation (\ref{qpdos}). 
We define the quasiparticle number $\tilde n$ as the integral of the sum of the spin up
and spin down quasiparticle density of states up to the Fermi level, 
\begin{equation}
\tilde n=\frac{2}{\sqrt{z_\uparrow z_\downarrow}}\!\!
\int\limits_{-\infty}^{0}\!\!\!
\frac{d\omega(\omega+\tilde\mu)}{\sqrt{(\omega+\tilde\mu)^2-\Delta\tilde\mu^2}}
 \rho_0\!\left(\!\frac{\sqrt{(\omega+\tilde\mu)^2-\Delta\tilde\mu^2}}{\sqrt{z_\uparrow
      z_\downarrow}}\right) .
\end{equation}
If we change the variable of integration to $\omega'$, where 
$$\omega'\sqrt{z_{\uparrow}z_{\downarrow}}=\sqrt{(\omega+\tilde\mu)^2-\Delta\tilde\mu^2},$$
the integration can be shown to be identical with that in equation
(\ref{ntotal}), using the fact that
$\mu_0=-\sqrt{\bar\mu_\uparrow\bar\mu_\downarrow}$. We then have an alternative
statement of Luttinger's theorem in the form $\tilde n=n$. This can also be
found by summing both spin components in (\ref{qpdeltaexc}), integrating over
$\omega$ and then converting the $\vk$-summation to an integral over the free
electron density of states $\rho_0(\omega)$. We can check in our numerical
results that the relation in this form holds. The occupation number $n$
can be calculated both from a direct evaluation of the number operator in
the ground state, and also by integrating the sum of the spectral densities
$\rho_{\sigma}(\omega)$ of the full local Green's function to the Fermi level. The value of  $\tilde n$ is
similarly determined from the integral over the total quasiparticle density of
states, $\tilde\rho_\sigma(\omega)$.
All three results were found to be in good agreement, to within one or
two percent deviation at the most.  

Before discussing the $\vk$-resolved spectra in detail we would like to ask
what the spectral weight $w_{\rm qp}$ of a quasiparticle excitation at
the Fermi level in the lower band is, such that the Green's function reads
there
\begin{equation}
  G_{\rm qp}(\omega)=\frac{w_{\rm qp}}{\omega-E^0_{\vkF,-}}.
\end{equation}
To calculate $w_{\rm qp}$, we can not focus on the spin dependent local
sublattice quantities, but have to sum over both sublattices or equivalently
the two spin components. The reason for this is that the antiferromagnetically ordered
state does not possess any net magnetization and has on average as many spin
up polarized as spin down electrons. The division in the $A$ and $B$
sublattices is convenient for the DMFT calculations but somewhat
artificial. In our case with hole doping the Fermi level lies within 
the lower band, which for the free quasiparticles is denoted by
$E^0_{\vk,-}$. The corresponding weight on the Fermi surface defined by
(\ref{ifs}) is then given by 
\begin{equation}
  w_{\rm  qp}=
\sum_{\sigma}z_{\sigma}u_{-}^{\sigma}(\epsilon_{\vkF})=\frac{z_{\uparrow}+z_{\downarrow}}2
 +\frac{(z_{\uparrow}-z_{\downarrow})\Delta\tilde\mu}{2|\tilde\mu|},
\label{qpweightafm}
\end{equation}
where the average of the renormalized chemical potential $\tilde\mu$ and the
difference $\Delta\tilde\mu$ were defined below equation (\ref{qpex}). From the
definition of $\Delta\tilde\mu$ we can see that the second term in
(\ref{qpweightafm}) is spin rotation invariant. The spectral quasiparticle weight
$w_{\rm  qp}$ on the Fermi surface depends not only on the renormalization
factors $z_{\sigma}$, but also on the renormalized chemical potentials $\tilde
\mu_{0,\sigma}$. The same result for the weight (\ref{qpweightafm})
can be obtained from the diagonal form of $\underline{G}_{\vk,\sigma}(\omega)$
and the spectral weight of the lower band. The weight $w_{\rm qp}$ corresponds
to the spectral weight $Z$ at the Fermi level as for example given in references
\cite{Dag94,STKHCCGMAKSYSM06,SGKCC06}. 
The first term of the result for $w_{\rm  qp}$ is like the arithmetic average of
$z_{\sigma}$.  From figures \ref{qpweightU3} and \ref{qpweightU6} we can see
that $z_{\uparrow}>z_{\downarrow}$ and from figure \ref{tmuU3}
that $\tilde \mu_{0,\downarrow}<\tilde \mu_{0,\uparrow}<0$. Therefore the
second term in (\ref{qpweightafm}) gives a positive contribution to the
spectral weight. At the end of the section in figure \ref{compweightU36} we
show values of $w_{\rm  qp}$ in comparison with the arithmetic average of
$z_{\sigma}$. 

In order to understand better the properties of the quasiparticle bands, we
now  compare the  quasiparticle spectrum with the  ${\vk}$-resolved spectral  
density $\rho_{{\vk},\sigma}(\omega)$  derived from the DMFT-NRG results.
\begin{figure}[!htb]
\begin{center}
\includegraphics[width=0.45\textwidth]{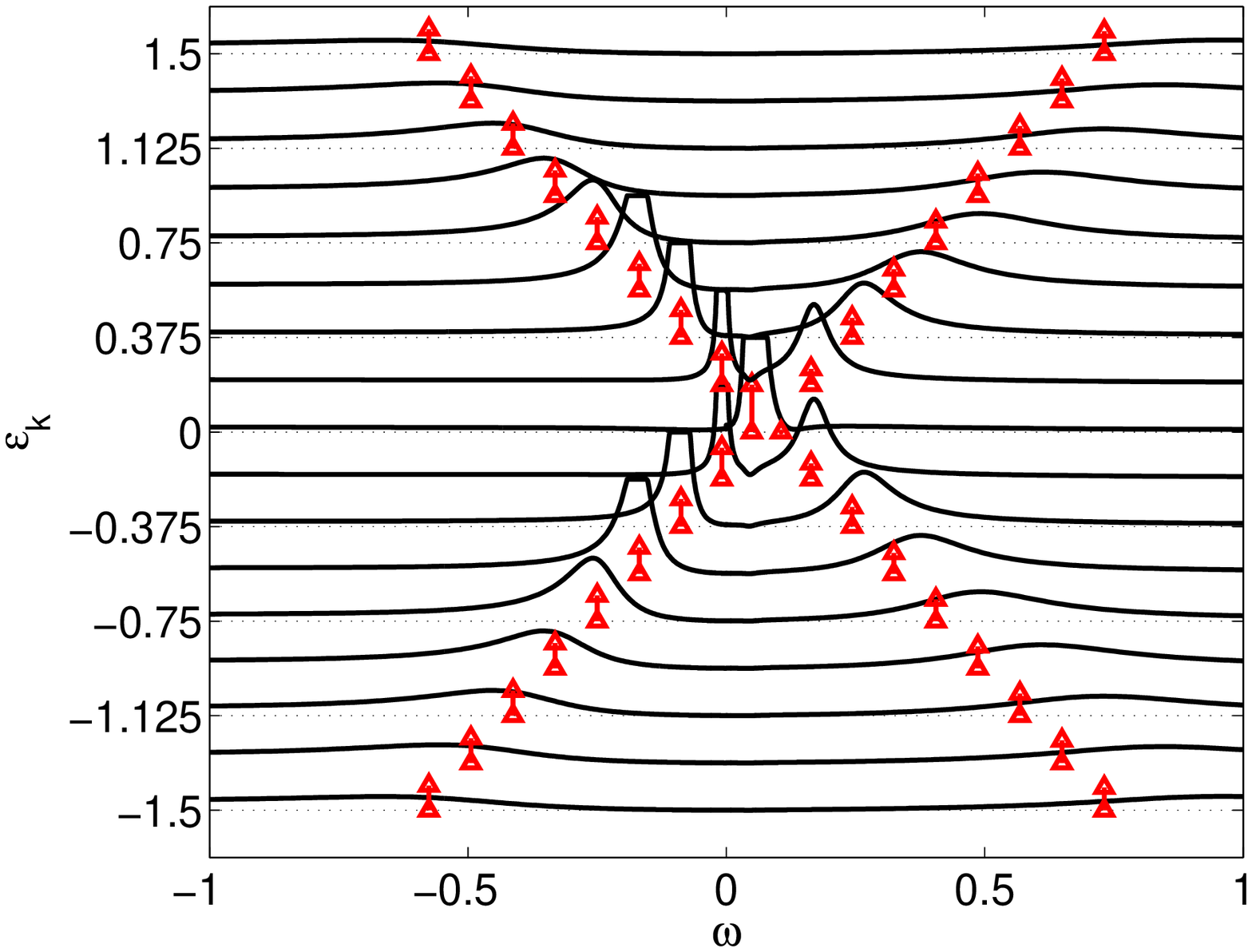}
\includegraphics[width=0.45\textwidth]{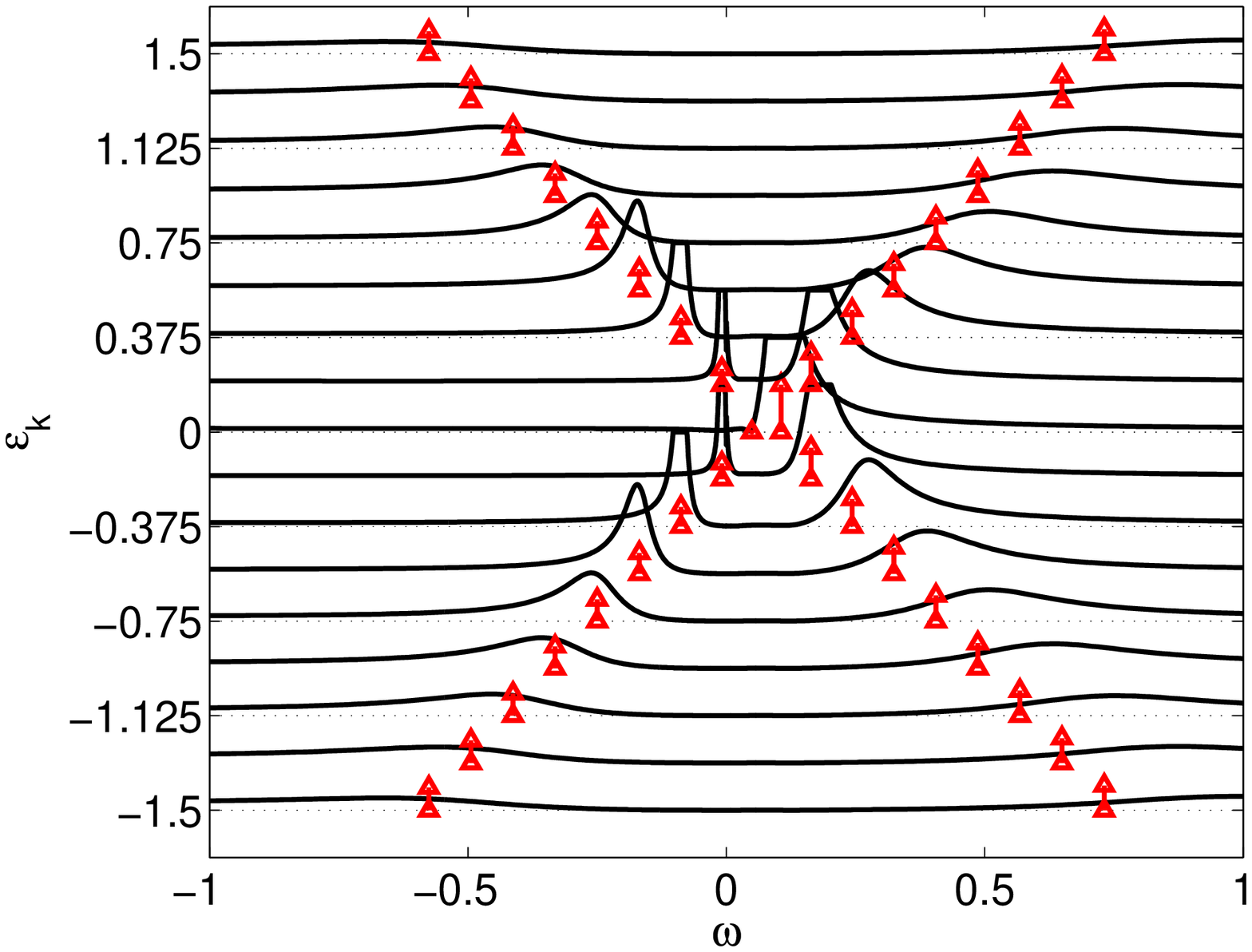}
\end{center} 
\vspace*{-0.5cm}
\caption{(Color online) The spectral density $\rho_{{\vk},\sigma}(\omega)$
for the spin-up electrons (upper panel) and spin-down (lower panel) plotted
as a function of $\omega$ and a sequence of values of $\epsilon_{\vk}$ for
$U=3$ and 12.5\% doping. Also shown with arrows are the positions of the
free quasiparticle excitations, with the height of the arrow indicating the
corresponding weight.}
\label{rhokomU3x0.9updown}
\end{figure}
\noindent
In figure  \ref{rhokomU3x0.9updown} we make a comparison for the case of  12.5\%
doping with $U=3$ for the Green's function  $G_{\vk,\sigma}(\omega)$ given in
equation (\ref{nkgf}),   
$\rho_{{\vk},\sigma}(\omega)=-{\rm Im}G_{\vk,\sigma}(\omega^+)/\pi$, where
$\omega^+=\omega+i\eta$, with $\eta\to 0$, with that derived for the free
quasiparticles,  $z_\sigma\tilde\rho^{(0)}_{{\vk},\sigma}(\omega)$ from
equation (\ref{qpdeltaexc}).
The delta-functions of the free quasiparticle results are indicated by arrows
with the height of the arrow indicating the value of the corresponding
spectral weight. The plots as a function of $\omega$ are shown for a sequence
values of  $\epsilon_{\vk}$ and, where the peaks in $\rho_{{\vk},\sigma}(\omega)$
get very narrow and high in the vicinity of the Fermi level, they have been truncated.
It  can be seen that the free quasiparticle results give a reasonable picture
of the form of  $\rho_{{\vk},\sigma}(\omega)$, particularly in the immediate
region of the Fermi level. There is considerable variation along the curves in
the way the overall spectral weight is distributed between the excitations
below and above the pseudo-gap as a function of $\epsilon_{\vk}$. This is most
marked in the region near the Fermi level for the spin-up electrons (upper
panel) where most of the spectral weight is in the lower band and it is much reduced in the
upper band, whereas the opposite is the case for the spin-down electrons.
This is reflected in the expression of the quasiparticle weights
$u_{\pm}^{\sigma}(\epsilon_{\vk})$ in equation (\ref{weights}). For instance,
$u_{-}^{\uparrow}(\epsilon_{\vk})$ corresponding to the lower band $E^0_{\vk,-}$
becomes maximal near the Fermi energy, whereas $u_{+}^{\uparrow}(\epsilon_{\vk})$
goes to zero there.
The finite width of the quasiparticle peaks in $\rho_{{\vk},\sigma}(\omega)$ 
can be described by a RPT, when we take into account the renormalized
self-energy $\tilde \Sigma_{\sigma}(\omega)$ in equation (\ref{rengfct}). If
we, for instance, use the the second order  approximation in $\tilde U$, which
was illustrated in the last section (\ref{afmrensig2}), we get a similar
behavior for small $\omega$ as seen for $\rho_{{\vk},\sigma}(\omega)$ in figure
\ref{rhokomU3x0.9updown}.   

\begin{figure}[!htbp]
\begin{center}
\includegraphics[width=0.45\textwidth]{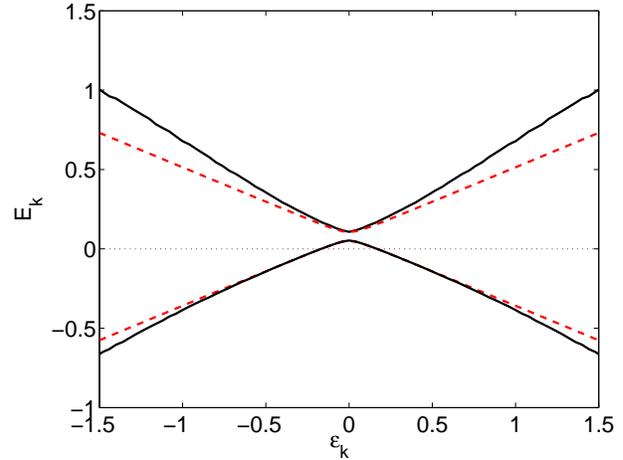}
\end{center} 
\vspace*{-0.5cm}
\caption{(Color online) A plot of the peak positions $E_{\vk,\pm}$ in the
  spectral density  $\rho_{{\vk},\sigma}(\omega)$ (full line) as a function of
  $\epsilon_{\vk}$ for $U=3$ and 12.5\%  doping compared with the free
  quasiparticle dispersion $E^0_{\vk}$  (dashed line).}
\label{qpexU3}
\end{figure}

From the positions of the peaks in the  $\rho_{{\vk},\sigma}(\omega)$
spectra we can deduce two branches of an effective dispersion $E_{\vk,\pm}$
for single particle excitations and compare it with the ones for the free
quasiparticles $E^0_{\vk,\pm}$. We give the results for $U=3$ in figure
\ref{qpexU3}. 

\noindent
It can be seen that $E^0_{\vk,-}$ tracks the peak in the lower
band  closely over a wide range of $\epsilon_{\vk}$, $-1.5<\epsilon_{\vk}<1.5$ (note
the bandwidth $W=4$). This is not the case in the upper band, where 
$E^0_{\vk,+}$ tracks the peak closely only in the lowest section that lies
closest to the Fermi level.
As one can see from the dotted line the Fermi level lies in the lower band and
intersects the lower band twice. This corresponds to the two values with
opposite sign $\epsilon^{\pm}_{{\vkF}}$ as can be see from equation (\ref{ifs}).

\begin{figure}[!htb]
\begin{center}
\includegraphics[width=0.45\textwidth]{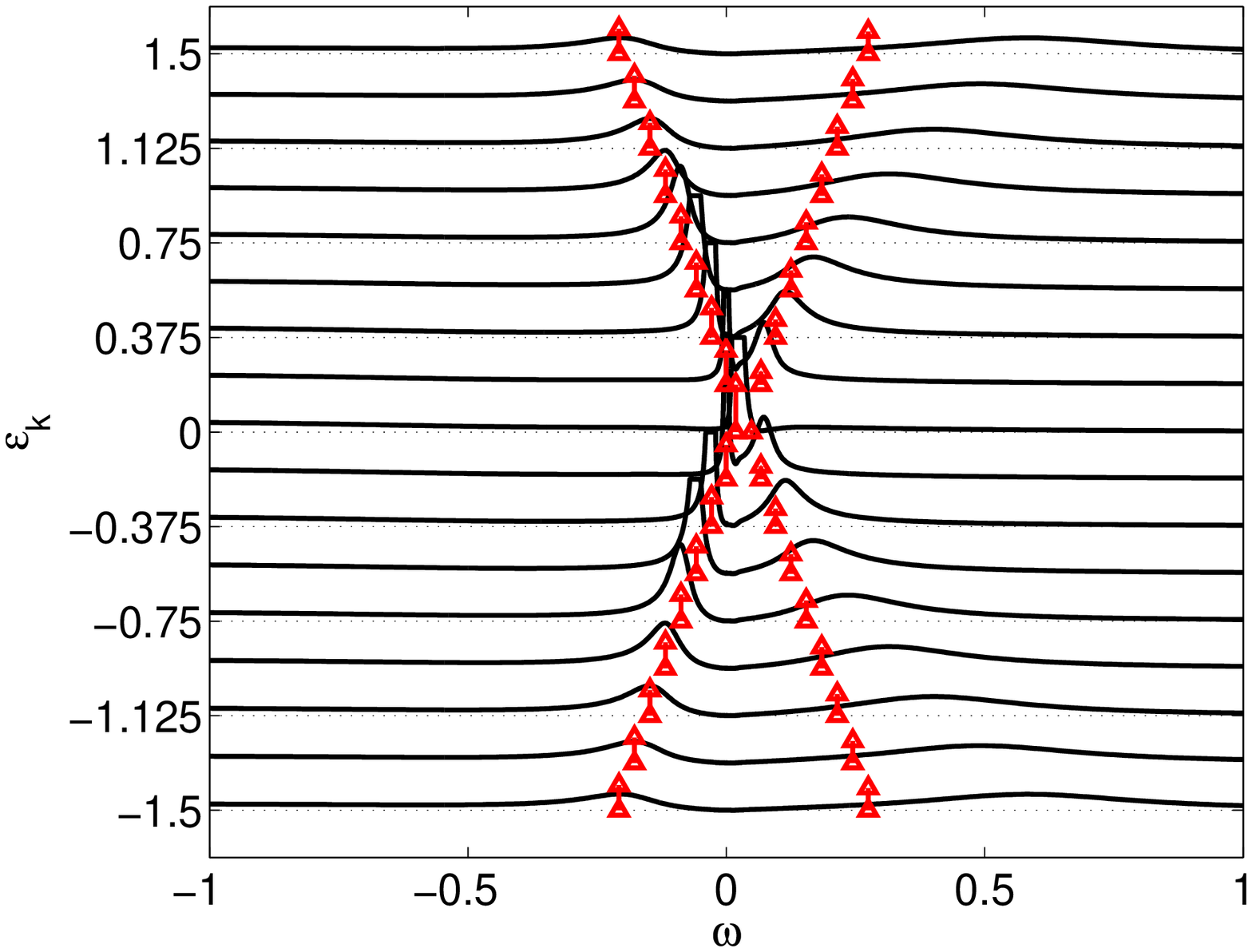}
\includegraphics[width=0.45\textwidth]{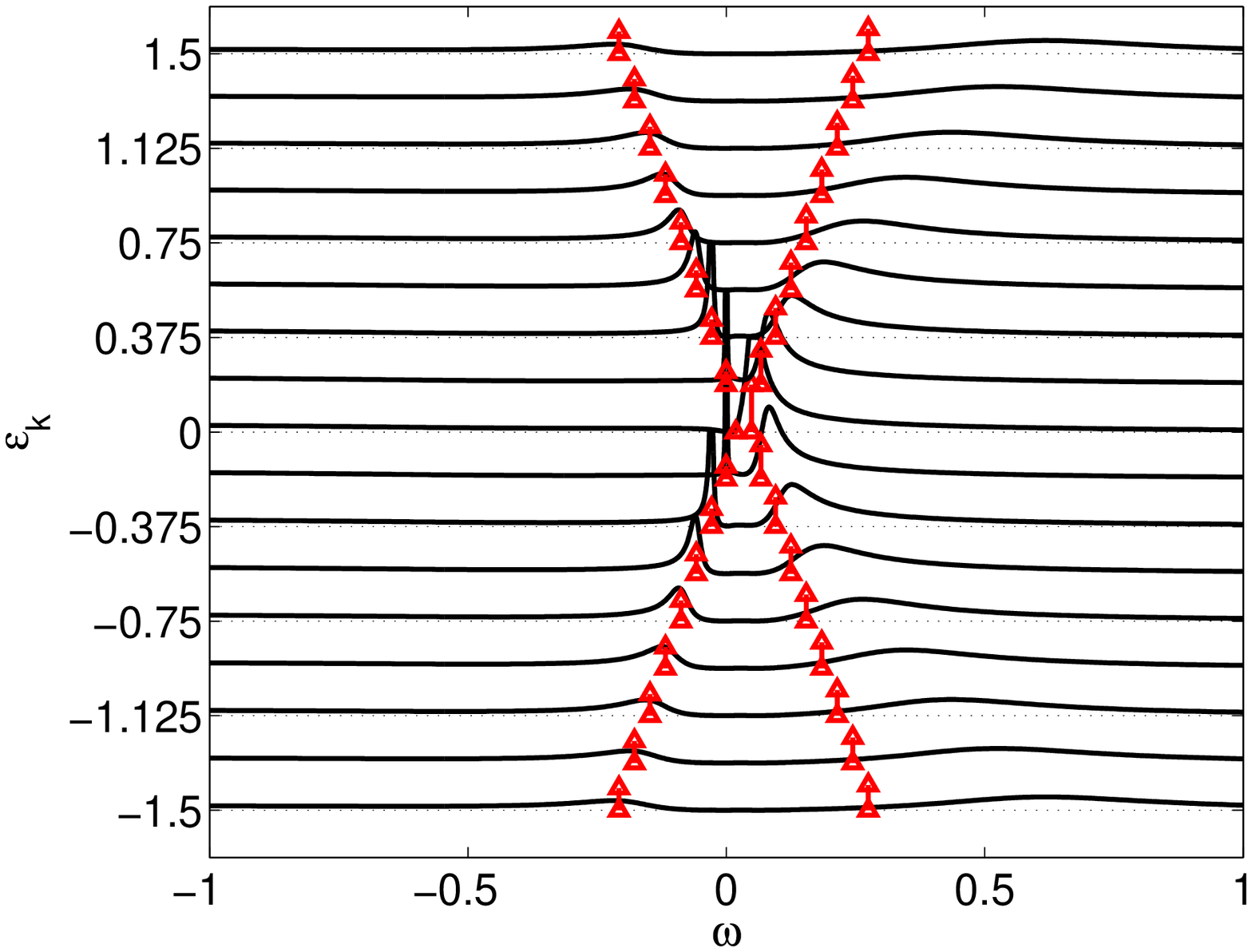}
\end{center} 
\vspace*{-0.5cm}
\caption{(Color online) The spectral density $\rho_{{\vk},\sigma}(\omega)$
for the spin-up electrons (upper panel) and spin-down (lower panel) plotted
as a function of $\omega$ and a sequence of values of $\epsilon_{\vk}$ for 
$U=6$ and 12.5\% doping. Also shown with arrows are the positions of the
free quasiparticle excitations, with the height of the arrow indicating the
corresponding weight.  }
\label{rhokomU6x0.9updown}
\end{figure}

The corresponding results for the $\vk$-resolved spectra for $U=6$ and also 12.5\%
 doping are shown in figure  \ref{rhokomU6x0.9updown}.

In order to compare well with the case $U=3$ we have chosen an identical range
for $\omega$ and $\epsilon_{\vk}$, although the large spectral peaks near the
energy are very close together in this presentation. 
It can be seen that the overall features are very  similar to those seen for 
$U=3$. For the spin up spectrum (upper panel) the peaks for the lower band have most of the
weight near the Fermi energy, whereas the upper band is suppressed there, and
vice versa for the opposite spin direction.
The lower bands are tracked well by the free quasiparticles, and we can see
that the  bands for the larger value of $U$ are significantly flatter.
This is also clearly visible in the following figure \ref{qpexU6}, where we
again compare the quasiparticle band with the peak position of the full
spectra.
On the range shown the lower band  $E_{\vk,-}$ completely coincides with the
free quasiparticle band  $E^0_{\vk,-}$.

\begin{figure}[!htbp]
\begin{center}
\includegraphics[width=0.45\textwidth]{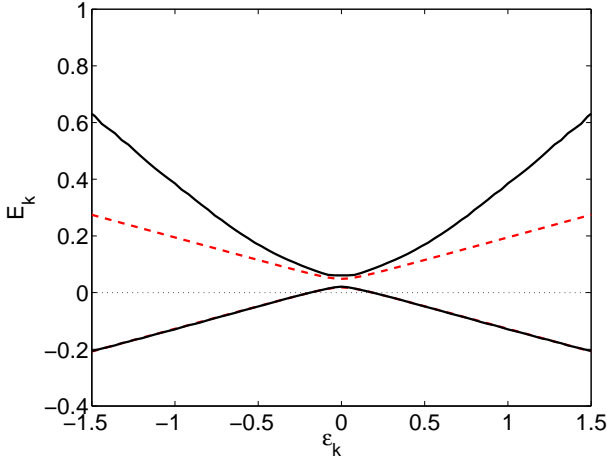}
\end{center} 
\vspace*{-0.5cm}
\caption{(Color online) A plot of the peak positions $E_{\vk,\pm}$ in the spectral density
  $\rho_{{\vk},\sigma}(\omega)$ (full line) as a function of $\epsilon_{\vk}$ for
  $U=6$ and 12.5\% doping compared with the free quasiparticle dispersion 
  $E^0_{\vk}$ (dashed line). On the range shown the lower band  $E_{\vk,-}$
  completely coincides with the free quasiparticle band  $E^0_{\vk,-}$.}
\label{qpexU6}
\end{figure}
\begin{figure}[!htbp]
\begin{center}
\includegraphics[width=0.45\textwidth]{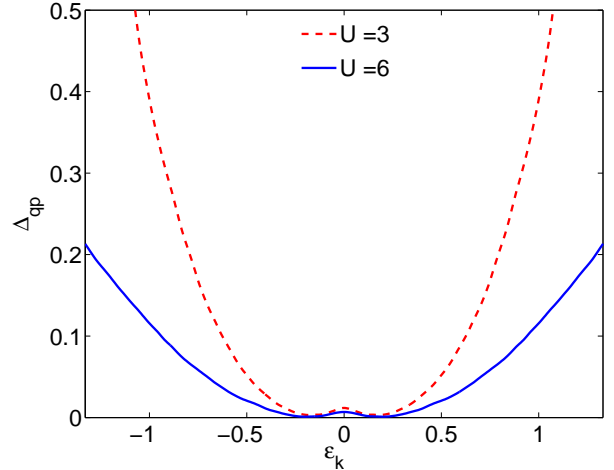}
\end{center} 
\vspace*{-0.5cm}
\caption{(Color online) A plot of the width of the peaks $\Delta_{\rm qp}$ in
  the spectral density of the majority spin $\rho_{{\vk},\uparrow}(\omega)$ as a function of
  $\epsilon_{\vk}$ for $U=3$ (dashed line) and $U=6$ (full line) and 12.5\%
  doping.} 
\label{qpwidthU36}
\end{figure}

\noindent
From the $\vk$-resolved spectra in figures \ref{rhokomU3x0.9updown} and
\ref{rhokomU6x0.9updown} we can also extract the width of the quasiparticle 
peak $\Delta_{\rm qp}$ in the spectral density $\rho_{{\vk},\sigma}(\omega)$
(majority spin $\sigma=\uparrow$). Its inverse $1/\Delta_{\rm qp}$ gives a
measure of the quasiparticle lifetime. The results for $\Delta_{\rm qp}$ for
the lower band $E_{\vk,-}$ for the two cases $U=3,6$ and 12.5\% doping are
shown in  figure \ref{qpwidthU36} as function of $\epsilon_{\vk}$. 

\noindent
This plot brings out more clearly the feature that can be seen already in figures
\ref{rhokomU3x0.9updown} and \ref{rhokomU6x0.9updown} (upper panel) that the
width increases sharply when we move away from the Fermi level and the values
for the width $\Delta_{\rm qp}$ for $U=6$ are significantly smaller than those
for $U=3$. This is in line with the fact that the local quasiparticle
interaction $\tilde U$ is smaller for the larger value of the bare interaction
$U$ as commented on earlier. The free quasiparticle picture is therefore even
more appropriate in the case with stronger interaction.
To numerical accuracy the width vanishes at $\epsilon^{\pm}_{{\vkF}}$ and
is finite for the interval
$\epsilon^{-}_{{\vkF}}<\epsilon_{\vk}<\epsilon^{+}_{{\vkF}}$ which lies within
the lower band but above the Fermi level. 

\begin{figure}[!htbp]
\begin{center}
\includegraphics[width=0.45\textwidth]{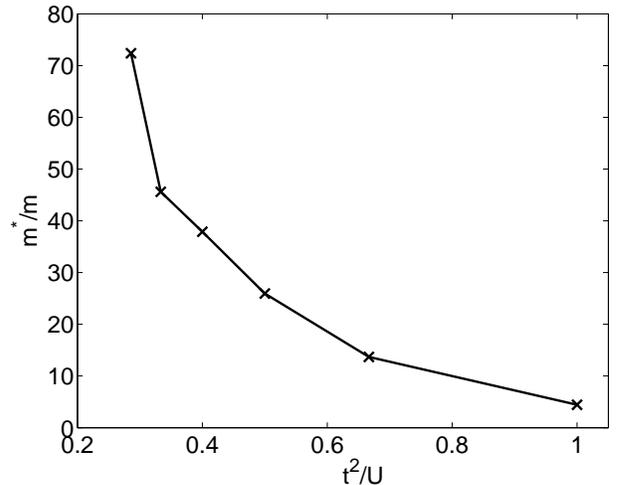}
\end{center} 
\vspace*{-0.5cm}
\caption{The ratio $m^*/m$ according to (\ref{effmassafm}) plotted over a
  range of $t^2/U$ for 7.5\% doping.}
\label{effmassx0.1}
\end{figure}

\begin{figure}[!htbp]
\begin{center}
\includegraphics[width=0.45\textwidth]{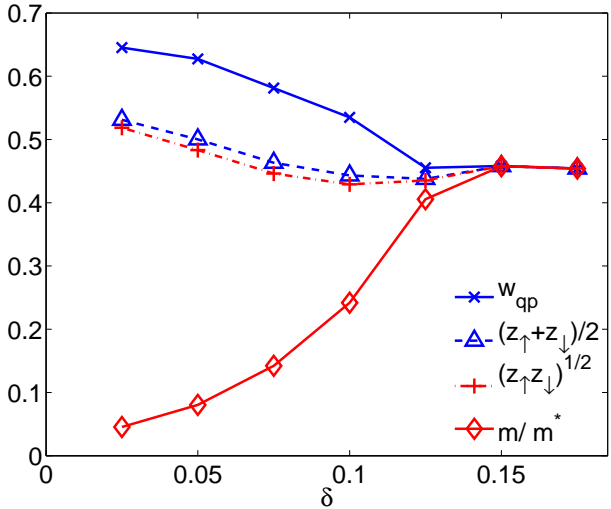}
\end{center} 
\vspace*{-0.5cm}
\caption{(Color online) Comparison of the spectral quasiparticle weight $w_{\rm qp}$
  from equation (\ref{qpweightafm}),  the arithmetic, $(z_{\uparrow}+z_{\downarrow})/2$, and
  geometric, $\sqrt{z_{\uparrow}z_{\downarrow}}$, average of the
  renormalization factors, and the inverse of the effective mass, $m/m^*$ from
  equation  (\ref{effmassafm}), for $U=3$  and a range of dopings.}
\label{compweightU36}
\end{figure}

Another quasiparticle property, the effective mass enhancement $m^*/m$, can be
extracted by calculating the derivative of $E^0_{\vk,-}$ in (\ref{qpex}) with
respect to $\epsilon_{\vk}$, which yields when evaluated at the Fermi energy
(\ref{ifs}), 
\begin{equation}
\frac{m^*}m=
\frac1{\sqrt{z_{\uparrow}z_{\downarrow}}}\frac{|\tilde \mu|}{\sqrt{\tilde
    \mu_{0,\uparrow}\tilde \mu_{0,\downarrow}}} .
\label{effmassafm}
\end{equation}
The effective mass enhancement therefore does not only depend on $z_{\sigma}$,
but also on the renormalized chemical potentials $\tilde \mu_{0,\sigma}$.
The general trend for $m^*/m$ as function of $U$ can be seen in figure
\ref{effmassx0.1} for the case of 7.5$\%$ doping. 

\noindent
The effective mass increases sharply for large $U$ as the hole motion
is energetically more costly in the ordered background. The fact that
the lower band for $U=6$  seen in figure \ref{qpexU6} is flatter 
than in the case $U=3$ in figure \ref{qpexU3} can be attributed  to
the larger effective mass.
We find a similar behavior of $m^*/m$ as function of $U$ for different
filling factors from the ones shown in figure \ref{qpwidthU36}. The
trend is that the effective mass enhancement is less pronounced for larger
doping, which is intuitively understandable by the quasiparticle motion in an
ordered background.
 
In the DMFT framework for the paramagnetic state as well as the case with
homogeneous magnetic field, the quasiparticle spectral weight $w_{\rm qp}$  and
the inverse of the effective mass enhancement $m/m^*$ can be described simply by the
renormalization factor $z_{\sigma}$. In figure \ref{compweightU36} we show a
comparison of the spectral quasiparticle weight $w_{\rm qp}$
(\ref{qpweightafm}), the arithmetic, $(z_{\uparrow}+z_{\downarrow})/2$, and
geometric, $\sqrt{z_{\uparrow}z_{\downarrow}}$, average of the renormalization
factors, and the inverse of the effective mass, $m/m^*$, (\ref{effmassafm})
for $U=3$ for various dopings. 

\noindent
As seen in this case with antiferromagnetic symmetry breaking these quantities
take a different form  (\ref{qpweightafm}) and (\ref{effmassafm}) and have
distinct values. For different values of $U$ the behaviour is qualitatively similar.
As a first approximation the quasiparticle spectral weight
$w_{\rm qp}$ corresponds to the arithmetic average of the renormalization
factors $z_{\sigma}$, whilst $m/m^*$ relates to the geometric average. In
general, one can, however, not omit the dependence on the renormalized chemical
potential as it gives a significant contribution as can be seen in figure
\ref{compweightU36}. This can be understood for example for the limit of zero
doping. The system then becomes an antiferromagnetically ordered  
insulator with spectral gap. The weights $z_{\sigma}$ tend to finite 
values, but the effective mass must diverge. This is found in equation 
(\ref{effmassafm}) since $\tilde\mu_{0,\uparrow}\to0$ for $\delta\to 0$,
and the trend can be seen in figure \ref{compweightU36}.

\section{Conclusions}

We have studied the field induced and spontaneous antiferromagnetic ordering
in the hole doped Hubbard model with DMFT-NRG calculations at $T=0$. A phase
diagram separating antiferromagnetic and paramagnetic solutions for different
values of doping and interactions $U$ ranging from zero to about 1.5 times the
bandwidth $W$ has been established and is in agreement with earlier results
by Zitzler et al. \cite{ZPB02}. Our main objective has been to analyze the
properties of the quasiparticle excitations in the metallic antiferromagnetic
state. We presented two different ways of calculating the parameters
$z_{\sigma}$ and $\tilde\mu_{0,\sigma}$, which define the renormalized
quasiparticles, and the two sets of results have been  shown to be in
agreement. We have also been able to deduce  the effective on-site
quasiparticle interaction $\tilde U$ from the NRG low lying excitations. 
The low energy properties of the local spectral function can be understood in
terms of the free quasiparticle picture. We have used the second order
perturbation expansion in powers of  $\tilde U$  to estimate the spectral
weight in the pseudo-gap region above the Fermi level. 

We have been able to compare the position of the  peaks  found in the
$\vk$-dependent spectral functions with the dispersion relation for the free
quasiparticles. The free quasiparticle dispersion gives a very good fit to the
position of these peaks in the lower band which intersects the Fermi level.
The quasiparticle lifetime, as deduced from the widths of the peaks in the
spectrum, increases for stronger interactions. This is consistent with
the fact that the on-site quasiparticle interaction $\tilde U$, which gives
the quasiparticles a finite lifetime, decreases with increase of $U$ in the
same range. We have also shown how the spectral quasiparticle weight at the
Fermi level $w_{\rm qp}$ and the effective mass can be deduced from the
parameters $z_{\sigma}$ and $\tilde\mu_{0,\sigma}$. The effective mass
is found to increase with the interaction, and it diverges in the limit of
zero doping whilst $w_{\rm qp}$ remains finite. 

We have found that Luttinger's theorem for the total electron density in the
antiferromagnetically ordered state holds within the numerical accuracy for the
range of dopings and interactions studied. This is a further indication that
many aspects of  Fermi liquid description may hold in situations  with
symmetry breaking.  

It is not easy to make a direct comparison of our results with earlier work
\cite{Dag94} analyzing the quasiparticle excitations in an metallic
antiferromagnet as these have been mainly  based on the $t-J$-model for one or
two holes in a finite cluster. However, at a semiquantitative level, 
the overall trend in our results seems to be similar to the results surveyed
by Dagotto, where the effective quasiparticle bandwidth $W_{\rm eff}$ is found
to decrease with decreasing $J$. This is line with our results if we identify
$W_{\rm eff}\sim m/m^*$ and $J\sim  t^2/U$ (see
figure \ref{effmassx0.1}). Our values for the spectral quasiparticle weight
$w_{\rm qp}$ are qualitatively similar to those presented as the wavefunction
renormalization $Z$ in the review article by Dagotto (see fig 27
\cite{Dag94}), and also the ones reported more recently \cite{SGKCC06}.  

\bigskip
\bigskip
\noindent{\bf Acknowledgment}\par
\bigskip
\noindent
We wish to thank N. Dupuis, D.M. Edwards, W. Koller, D. Meyer and  A. Oguri
for helpful discussions and  W. Koller and D. Meyer for their contributions to
the development of the NRG programs. We also acknowledge stimulating
discussions with G. Sangiovanni.
One of us (J.B.) thanks the Gottlieb Daimler and Karl Benz Foundation, the
German Academic exchange service (DAAD) and the EPSRC for financial support.

\bibliographystyle{prsty}
\bibliography{artikel,biblio1}

\end{document}